\documentclass[12pt]{article}
\usepackage{amsmath,amsthm,amssymb}    
\usepackage{paralist}
\usepackage{graphics}   
\usepackage{verbatim}   
\usepackage{subfigure}  
\usepackage{hyperref}   
\usepackage{graphicx,epsfig}
\usepackage{multirow}
\usepackage{float}
\usepackage{color}

\restylefloat{table}

\makeatletter
 
 \@addtoreset{equation}{section}
\makeatother

\newtheorem{Theorem}{Theorem}[section]
\newtheorem{Lemma}[Theorem]{Lemma}
\newtheorem{Proposition}[Theorem]{Proposition}
\newtheorem{Prop}[Theorem]{Proposition}
\newtheorem{Remark}[Theorem]{Remark}

\begin{document}

\title{Coexisting Stable Equilibria in a Multiple-allele Population
Genetics Model}
\author{Linlin Su\footnote{Email:\,\,lsu@wpi.edu}\quad Colbert
Sesanker\footnote{Email colbertsesanker@wpi.edu}\quad
and\,\,\,\,Roger Lui\footnote{Email:\,\,rlui@wpi.edu} \\
Department of Mathematical Sciences \\
Worcester Polytechnic Institute\\
100 Institute Road\\
Worcester, MA 01609}
\date{}

\maketitle

\begin{abstract}
In this paper we find and classify all patterns for a single locus
three- and four-allele population genetics models in continuous
time. A pattern for a $k$-allele model means all coexisting locally
stable equilibria with respect to the flow defined by the equations
$\dot{p}_i = p_i(r_i-r), i=1,\dots,k,$ where $p_i, r_i$ are the
frequency and marginal fitness of allele $A_i$, respectively, and
$r$ is the mean fitness of the population. It is well known that for
the two-allele model there are only three patterns depending on the
relative fitness between the homozygotes and the heterozygote. It
turns out that for the three-allele model there are $14$ patterns
and for the four-allele model there are $117$ patterns. With the
help of computer simulations, we find $2351$ patterns for the
five-allele model.  For the six-allele model, there are more than
$60,000$ patterns. In addition, for each pattern of the three-allele
model, we also determine the asymptotic behavior of solutions of the
above system of equations as $t \rightarrow \infty$. The problem of
finding patterns has been studied in the past and it is an important
problem because the results can be used to predict the long-term
genetic makeup of a population.
\end{abstract}
\bigskip
\noindent Key words:\,\, k-allele model, stable equilibrium, patterns,
computer simulations.\smallskip




\section{Introduction}\label{intro}
Selection is an important driving force in evolution and is due
mainly to the differences in fitness to survival between genotypes
in a population. Since the 1920's many mathematical models for
selection have been developed and studied and many results are known
\cite{Nagylaki,Burger}. In this paper we give a partial answer to
an old but still open problem mentioned in \cite[p.38]{Burger}
concerning the maximum possible number of coexisting stable
equilibria in the evolution of gene frequencies under selection
forces in a single-locus multiple alleles population genetics
model.\smallskip

Consider a diploid population whose members possess a gene that
occurs in $k$ different forms, called alleles, located at an
autosomal locus.  Let the fitness of genotype $A_iA_j$ be denoted by
$r_{ij}$ and $R = (r_{ij})$ is the $k$ by $k$ symmetric fitness
matrix. Let $p_{i}$ be the frequency of allele $A_i$. Then, assuming random mating and discrete-time
non-overlapping generations, the frequency of allele $A_i$
in the next generation is given by
\begin{equation*}
p_i^{'} = p_ir_i/r\,,\quad\quad i=1,\dots,k\,,
\end{equation*}
where $r_i = \sum_j r_{ij}p_j$ is the marginal fitness of allele
$A_i$ and $r = \sum_{i} r_ip_i$ is the mean fitness of the
population. For continuous-time overlapping generations, above
equations are replaced by
\begin{equation}\label{basicmodel}
\dot{p}_i = p_i(r_i - r)\,,\quad\quad i=1,\dots,k\,,
\end{equation}
which is only an approximation, derived under the assumption of
Hardy-Weinberg equilibrium \cite[Section 10.1]{Burger}. In this
paper, we focus our study on the continuous-time model
(\ref{basicmodel}). Our main idea is to determine the signs of
eigenvalues of the Jacobian matrix evaluated at an existing
equilibrium to determine if the equilibrium is stable or not. The
definitions of mathematical terms used in this paper will be given
in Section~\ref{term}.
\smallskip

Let $p_i(t), i=1,\dots,k$ be the solutions of (\ref{basicmodel}).
Since they are the frequencies at the same locus, they must add up
to one; that is $\sum p_i(t) = 1$. Therefore, (\ref{basicmodel}) is
actually a system of $(k-1)$ equations with $p_k$ replaced by
$(1-p_1-\dots-p_{k-1})$. By relabeling the alleles, we may assume
that $r_{ii}$'s are non-increasing in $i$ but for simplicity, we
assume that they are decreasing; that is,
\begin{equation}\label{riicond}
r_{11} > r_{22} > \dots > r_{kk}\,.
\end{equation}

This paper is organized as follows. In Section~\ref{term}, we
present the mathematical background necessary to understand this
paper. In Section~\ref{2allele}, we present the well known results
of the two-allele model, define what is a pattern, and review the
literature on the subject of finding patterns. In
Section~\ref{generic}, we explain how one can obtain the eigenvalues
of the Jacobian matrix evaluated at an equilibrium. In
Section~\ref{3allele}, we give a rigorous proof that there are $14$
patterns for the three-allele model. A complete description of the
dynamics of the solutions of \eqref{basicmodel} is also included. In
Section~\ref{higherallele}, we present our computer simulations that
led to the discovery of patterns for the four- and five-allele
models. Section \ref{conclusion} is conclusion. In order to make
this paper accessible to a wider audience, we omit most of the
proofs and instead focus on examples to illustrate our results.

\section{Mathematical Background}\label{term}
\setcounter{equation}{0}

In this section, we give a brief introduction to the definitions of
the mathematical terms used in this paper. Further details may be
found in the book \cite{leah}. Let
\begin{equation}\label{system}
f_i(p_1,\dots,p_{k-1}) = p_i(r_i - r),\quad\quad i=1,\dots,k-1,
\end{equation}
where $r_i, r$ are defined in section~\ref{intro}  with $p_k = 1-
p_1 - \dots - p_{k-1}$. An \emph{equilibrium} solution, ${\bf p}^* =
(p_1^*,\dots,p_{k-1}^*)$, is a constant solution of the nonlinear
system of equations (\ref{system}). The \emph{Jacobian} matrix of
$f_i$'s is
\[
J = (J_{ij})\quad\mbox{where the $i^{th}$ row $j^{th}$ column component is}\quad
J_{ij} = \displaystyle\frac{\partial f_i}{\partial p_j}\,.
\]
If we want to find the behavior of the solutions of
(\ref{basicmodel}) as time goes to infinity ($t \rightarrow \infty$)
with a starting point near an equilibrium ${\bf p}^*$, one can
instead study the linear system
\begin{equation}\label{linear}
\dot{{\bf p}} = J^*\, {\bf p}\,,
\end{equation}
where $J^*$ is the constant matrix obtained by evaluating $J$ at the
equilibrium ${\bf p}^*$. For further details and examples see
\cite[Section 5.7]{leah}.\smallskip

The behavior of solutions to (\ref{linear}) depends on the signs of
the eigenvalues of $J^*$. A constant $\lambda$, which may be a
complex number, is an eigenvalue of $J^*$ if there exists a nonzero
$k-1$-dimensional vector, ${\bf x}$, such that $J^*{\bf x} = \lambda
{\bf x}$. It is well known that $J^*$ has $k-1$ eigenvalues and
${\bf p}^*$ is \emph{asymptotically stable} if and only if all the
eigenvalues of $J^*$ have negative real-parts. If ${\bf p}^*$ is
asymptotically stable, then there exists a small ball centered at
${\bf p}^*$ (also called a neighborhood) such that the solutions of
(\ref{basicmodel}) with initial value inside this ball converge to
${\bf p}^*$ as $t \rightarrow \infty$. In this paper, for
simplicity, we say ${\bf p}^*$ is stable instead of asymptotically
stable. This should not cause any confusion. It turns out that for
the population genetics model, the eigenvalues of $J^*$ is always
real and a large part of section~\ref{generic} is to show how to
find the signs of the eigenvalues of $J^*$.  There are some linear
algebra terms needed to understand this paper. Details may be found
in the book \cite{Strang}. \emph{Determinant} of $J^*$ is the
product of all the eigenvalues of $J^*$ and the \emph{trace} of
$J^*$ is the sum of all the eigenvalues of $J^*$. If $J^*$ is of
size $2$ (three-allele case), let $\lambda_1, \lambda_2$ be the
eigenvalues of $J^*$, then $\mbox{det}(J) = \lambda_1\lambda_2$ and
$\mbox{tr}(J) = \lambda_1 + \lambda_2$. Therefore, we can determine
the signs of the eigenvalues from the signs of $\mbox{det}(J^*)$ and
$\mbox{tr}(J^*)$. This fact will be used in Section~\ref{3allele}.
Let $A$ be a square matrix. Choose any entry $a_{ij}$ of $A$ that
lies in the $i^{th}$-th row and $j^{th}$-column of $A$. Cross out
the $i^{th}$-row and $j^{th}$-column of $A$ and denote the
determinant of the resulting matrix by $M_{ij}$. Then
$(-1)^{i+1}M_{ij}$ is called the \emph{cofactor} of $a_{ij}$ and $M
= (M_{ij})$ is called the cofactor matrix of $A$. Cofactors will be
used in Section~\ref{generic}. The transpose of the cofactor matrix
$M$ is called the \emph{adjoint} of $A$. A \emph{permutation matrix}
is a matrix containing only $1$'s and $0$'s with each row and column
containing exactly one $1$. Multiplying $A$ on the left and right by
the same permutation matrix will interchange certain rows and
columns $A$.
\smallskip

Let us return to the study of our nonlinear system (\ref{basicmodel}).
Since the frequencies of the alleles always lie between $0$ and $1$
and add up to one, the solutions of (\ref{basicmodel}) always lie
inside the \emph{invariant set}
\begin{equation}\label{deltadef}\Delta =
\{(p_1,\dots,p_{k-1})\,:\,0 \leq p_i \leq
1\;\mbox{for}\;i=1,\dots,k-1,\;\sum_{i=1}^{k-1} p_i \leq 1\}\,.
\end{equation}
Invariant here means solutions of (\ref{basicmodel}) that start
inside $\Delta$ remain in $\Delta$ for all future time.  Also,
because $\Delta$ is invariant, all equilibria of (\ref{basicmodel})
lie inside $\Delta$. A \emph{boundary} equilibrium, ${\bf p}^*$, is
an equilibrium where at least one of the components of ${\bf p}^*$
is zero. An \emph{interior} equilibrium is one where all the
components are positive. It is known that solutions to
(\ref{basicmodel}) always converge to a stable equilibrium as $t
\rightarrow \infty$ \cite[p. 38]{Burger}. If there are more than
one stable equilibrium in $\Delta$, then $\Delta$ can be partitioned
into non-overlapping regions by hypersurfaces (which are curves when
$k=3$)  called \emph{separatrices}. Each of these regions contains
one stable equilibrium and solutions starting in a region will
converge to the stable equilibrium in that region as $t \rightarrow
\infty$. As an example, consider the first figure in
Figure~\ref{fig1} in the proof of Theorem~\ref{3allelemain}. The
vertices $(1,0)$ and $ (0,1)$ are the only stable equilibria. The
separatrix is the curve shown joining the origin and $P_3$ to the
interior equilibrium. Solutions that start on the separatrix
converge to the interior equilibrium while solutions that start
above or below it converge to $(0,1)$ or $(1,0)$, respectively as $t
\rightarrow \infty$.

\section{Two-allele Model and Finding All Possible Patterns}\label{2allele}
\setcounter{equation}{0}

Let us consider the simplest case with two alleles. Since
$p_1+p_2=1$, $\Delta = [0,1]$ and system (\ref{basicmodel}) may be
reduced to the single equation
\begin{equation}\label{twoallele}
\dot{p}_1=p_1(1-p_1)(\Delta_{12}p_1+r_{12}-r_{22})\,,
\end{equation}
where $\Delta_{12}=r_{11}+r_{22}-2r_{12}$. The following result is
well known \cite{Nagylaki}.
\begin{Prop}\label{2alleleprop}
For the two-allele model (\ref{twoallele}) with $r_{11} > r_{22}$,
$p_1=0$ and $p_1=1$ are both equilibrium and a third interior
equilibrium $p^* = (r_{22}-r_{12})/\Delta_{12}$ exists if and only
if $r_{12}>r_{11}$ or $r_{12}<r_{22}$. There are three cases: (i)
$r_{22}<r_{12}<r_{11}$ : $1$ is stable, $0$ is unstable and $p^*$
does not exist; (ii) $r_{12}>r_{11}$ : $0, 1$ are both unstable and
$p^{\ast}$ exists and is stable; (iii) $r_{12}<r_{22}$ : $0, 1$ are
both stable, $p^{\ast}$ exists and is unstable.  We call theses
three cases the (heterozygote) intermediate, superior, and inferior
cases, respectively. If one of the alleles is completely dominant;
i.e. $r_{12} = r_{11}$ or $r_{12} = r_{22}$, then $p^*$ does not
exist and (\ref{twoallele}) reduces to the intermediate case with a
double zero either at $p_1=0$ or at $p_1=1$.
\end{Prop}

The above proposition implies that for the two-allele model there
are only three patterns depending on the values of $r_{12}$ relative
to $r_{11}$ and $r_{22}$. Pattern 1 is when $1$ is the only stable
equilibrium, pattern 2 is when $p^*$ is the only stable equilibrium,
and pattern 3 is when $0$ and $1$ are both stable and the interior
equilibrium $p^*$ is unstable.  In Pattern 3, $0$ and $1$ are
coexisting stable equilibria. We denote these three patterns by
$\{1\}, \{p^*\}$, and $\{0,1\}$, respectively. From this example, we
shall define \emph{pattern} as the number and locations of all
coexisting stable equilibria.  Note that $\{1\}$ and $\{p^*\}$ are
different patterns because $1$ is a boundary equilibrium but $p^*$
is an interior equilibrium. However, even though $p^*$ varies
depending on the value of $r_{12}$, we still consider $\{p^*\}$ as
one pattern because it lies in the interior. For the $k$-allele
model there are $(2^k-1)$ possible equilibria but the number of
coexisting stable equilibria in any pattern is considerably less.
For example, for the five-allele model there are $2^5 - 1 = 31$
possible equilibria but no pattern can contain more than $6$ stable
equilibria.\smallskip

The problem of finding all patterns for a $k$-allele model has been
studied extensively. In a series of papers
\cite{broom,vickers1,vickers2,vickers4,vickers3}, the authors
studied \emph{maximal patterns}, that is, patterns which are not a
subset of another pattern, for $3,4,5$ alleles. Cannings and Vickers
conjectured that any subset of a pattern is also a pattern
\cite{vickers1,vickers4}. Thus, if this conjecture is true, one may
specify the complete set of patterns by specifying the complete set
of maximal patterns. This is the approach adopted in the literature.
In this paper, we give a complete list of patterns for the case
$k=3$ (proved analytically), $k=4$ (proved and confirmed by computer
simulations), and $k=5$ (with a few unsolved cases) without assuming
that the conjecture is true. Note also that \eqref{riicond} is not
posited in the literature. Therefore, patterns that are rotations of
each other can be obtained by multiplying the fitness matrix $R$ on
both sides by a permutation matrix and thus are considered as the
same pattern. However, under condition \eqref{riicond}, rotation of
a pattern is not necessarily a pattern. For example, for the
three-allele model, $\Delta$ is the triangle with vertices at
$(1,0), (0,1)$ and $(0,0)$. The vertex $\{(1,0)\}$ is a pattern but
the vertices $\{(0,1)\}$ and $\{(0,0)\}$ are not patterns even
though they can be obtained from $\{(1,0)\}$ by rotation. Therefore,
we treat patterns that are rotations of each other as different
patterns. For this reason, the number of patterns we obtained in
this paper is greater than those obtained in the literature. The
method we used in this paper, counting signs of eigenvalues of the
Jacobian matrix evaluated at the existing equilibria, is also
completely different from the method used in the literature.
Furthermore, we also consider the case of complete dominance and
describe all possible asymptotic behavior of the solutions of
(\ref{basicmodel}) when $k=3$.\smallskip

\section{Properties of System (\ref{basicmodel})}\label{generic}
\setcounter{equation}{0} We begin with the problem of existence of
interior equilibria.  Let $R_{ij}$ be the cofactor of $r_{ij}$ in
the fitness matrix $R$ and let $U_{i}=\sum_{j}R_{\,ij}$. The
following result is due to Mandel \cite{Mandel}.

\begin{Proposition}\label{interiorformula} Suppose $\det R\neq 0$.
An interior equilibrium exists if and only if all $U_i$'s are
non-zero and have the same sign. Moreover, if the interior
equilibrium exists, it is given by
\[p_i = \frac{U_{i}}{\sum_{j}U_{j}}\,,\quad i=1,\dots,k\,.\]
\end{Proposition}

The degenerate case $\det R=0$ is considered in \cite{HuSe} and
\cite{Heiden}. In this paper, we always assume that the determinant
of $R$ and all its submatrices are non-zero.  A submatrix of $R$ is
the matrix obtained by erasing any number of rows and the
corresponding columns of $R$.
\smallskip

Next we consider the stability of equilibria of system
\eqref{basicmodel}. It is well known that the mean fitness $r$,
defined in Section~\ref{intro}, is strictly increasing along
non-constant solutions of \eqref{basicmodel} \cite{Kingman2}.
Therefore, $r$ is what is commonly called a \emph{Lyapunov function}
and the stable equilibria are identified with the local maxima of
$r$. Quite a few necessary and sufficient conditions for the
stability of an existing equilibrium can be obtained by studying the
properties of local maxima of $r$, see \cite{Mandel2,HuSe} and the
references therein. Another more direct approach is to investigate
the eigenvalues of the Jacobian matrix $J$ evaluated at an
equilibrium of system \eqref{basicmodel}. It can be shown that all
such eigenvalues are real. A proof of this fact for the
discrete-time model is given in \cite[Section 3]{Kingman}. We adopt
the second more direct approach in this paper.\smallskip

Let ${\bf p}^*=(p_1^*, \ldots, p_k^*)$ be an equilibrium whose
components add up to one. Let $K$ be the set $\{1, \ldots, k\}$ and
define $L=\{i\in K: p_{i}^*>0\}$ and $\bar{L}=\{i\in K:
p_{i}^*=0\}$. The following proposition gives the formula of the
Jacobian matrix $J$ which will be used extensively in this paper.
The result is probably known but we are unable to find a reference
for it so we present a proof here. Readers not interested in the
proof may simply read Remark~{\ref{remark1}} to understand how this
proposition is used.
\smallskip

\begin{Proposition}\label{Jacobian}Let $l \in K$ and replace $p_l$ by $(1-\sum_{i\neq l}p_{i})$ in
(\ref{basicmodel}). Then the $i^{th}$-row, $j^{th}$-column of the
Jacobian matrix $J$ of \eqref{basicmodel} (as a system of $(k-1)$
equations) evaluated at an equilibrium point ${\bf p}^*$ is given by
\begin{equation*}
J_{ij} = \delta_{ij}(r_{\hat{i}}-r)|_{{\bf
p}^*}+p_{\hat{i}}^*[(r_{\hat{i}\hat{j}} -
r_{\hat{i}l})-2(r_{\hat{j}}-r_l)|_{{\bf p}^*}]\,,\quad
i,j=1,\dots,k-1\,,
\end{equation*}
where $\hat{i} = i$ if $1\leq i<l$, $\hat{i} = i+1$ if $i\geq l$;
namely,
\begin{equation*}
J_{ij} = \left\{
\begin{array}{ll}
p_{\hat{i}}^{*}[(r_{\hat{i}\hat{j}} -
r_{\hat{i}l})-2(r_{\hat{j}}-r_l)|_{{\bf p}^*}]
&\mbox{if}\quad\hat{i}\in L\,,\\
\delta_{ij}(r_{\hat i}-r)|_{{\bf p}^*} &\textrm{if}\quad\hat{i}\in
\bar{L}\,.
\end{array}\right.
\end{equation*}
In particularly, if ${\bf p}^*$ is an interior equilibrium so that
$\bar{L} = \emptyset$, then
\[
J_{ij} = p_{\hat{i}}^{*}(r_{\hat{i}\hat{j}} - r_{\hat{i}l})\,,\quad
i,j=1,\dots,k-1\,.
\]
\end{Proposition}
\begin{proof}
$J_{ij} = \frac{\partial}{\partial
p_{\hat{j}}}\,[p_{\hat{i}}(r_{\hat{i}} - r)]|_{{\bf p}^*} =
\delta_{\hat{i}\hat{j}}(r_{\hat{i}}-r)|_{{\bf
p}^*}+p_{\hat{i}}^*\,\frac{\partial}{\partial
p_{\hat{j}}}(r_{\hat{i}} - r)|_{{\bf p}^*}$. Now $r_{\hat{i}} =
\displaystyle\sum_{\ell \neq l} r_{\hat{i}\ell} p_{\ell} +
r_{\hat{i}l}(1-\sum_{\ell \neq l}p_{\ell})\,.$  Therefore,
$(r_{\hat{i}})_{p_{\hat{j}}} = r_{\hat{i}\hat{j}} - r_{\hat{i}l}$.
On the other hand, $r = \displaystyle\sum_{\ell \neq l}r_{\ell}
p_{\ell} + r_l(1-\sum_{\ell \neq l}p_{\ell})\,.$  Therefore,
\begin{eqnarray*}
r_{p_{\hat{j}}} &=& \sum_{\ell \neq l} (r_{\ell})_{p_{\hat{j}}}
p_{\ell} + r_{\hat{j}} +
(r_l)_{p_{\hat{j}}} p_l - r_l\\
&=& \sum_{\ell \neq l} (r_{\ell \hat{j}} - r_{\ell l})p_{\ell} +
r_{\hat{j}} +
\left(\sum_{\ell=1}^k r_{l\ell} p_{\ell}\right)_{p_{\hat{j}}} p_l - r_l \\
&=& r_{\hat{j}} - r_l - r_{\hat{j}l}p_l  + r_{ll}p_l + r_{\hat{j}} +
r_{l\hat{j}}p_l - r_{ll}p_l - r_l = 2(r_{\hat{j}}-r_l).
\end{eqnarray*}
Noting that $r_i=r$ for $i\in L$ and at an interior equilibrium,
$r_1 =\dots = r_k$, the conclusion of the proposition follows.
\end{proof}

It can also be shown that at an interior equilibrium ${\bf p}^*$ the determinant of the Jacobian matrix $J$ is given by
\begin{equation}\label{detJ}
\mbox{det}(J) =p_{1}^{*}p_{2}^{*}\cdots p_{k}^{*}(U_{1}+U_{2}+\cdots+U_{k}).
\end{equation}

\begin{Remark}\label{remark1}\,\,{\rm If we choose $l$ to be in $L$, then
\begin{equation*}
J_{ij} = \left\{\begin{array}{ll}p_{\hat i}^{*}(r_{\hat{i}\hat{j}} - r_{\hat{i}l})\quad & \mbox{if}\quad\hat{i},\hat{j}\in L\,,\\
\delta_{ij}(r_{\hat i}-r)|_{{\bf p}^*} \quad&
\textrm{if}\quad\hat{i}\in \bar{L}\,.
\end{array}\right.
\end{equation*}
This implies that the eigenvalues of $J$ consist of $(r_i-r)|_{{\bf
p}^*}$ ($i\in \bar{L}$) and the eigenvalues of the matrix
$(p_i^*(r_{ij} - r_{il}))_{i,j\in L-\{l\}}$. This formula greatly
simplifies computation of the eigenvalues of $J$. For example, for a
five-allele model, if we want to find the eigenvalues of $J$ at the
boundary equilibrium on the side where $p_2 = 0, p_5 =0$, all we
have to do is find the interior equilibrium of the three-allele
model with fitness matrix obtained by eliminating the second and
fifth rows and columns of $R$. Let this interior equilibrium be
denoted by $(\bar{p}_1, \bar{p}_2, \bar{p}_3)$ and let $p^* =
(\bar{p}_1,0,\bar{p}_2,\bar{p}_3,0)$. Then according to the above
proposition with $l = 4$, the eigenvalues of $J$ evaluated at $p^*$
consists of the eigenvalues of the matrix
\[
\left(
\begin{array}{cc}
\bar{p}_1(r_{11} - r_{14}) &\bar{p}_1(r_{13} - r_{14})\\
\bar{p}_2(r_{31} - r_{34}) &\bar{p}_2(r_{33} - r_{34})
\end{array}
\right)
\]
and ${\bf r}_i\cdot{\bf p}^* - R\,{\bf p}^*\cdot{\bf p}^*,\, i=2, 5$
where ${\bf r}_i$ is the $i^{th}$-row of the matrix $R$.}
\end{Remark}

\begin{Remark}\label{remark3}
{\rm From Section~\ref{term}, an equilibrium is stable if all the
eigenvalues of $J$ at that equilibrium have negative real parts and
unstable if at least one eigenvalue has positive real part.
Moreover, Equation (14) in \cite{Kingman} shows that zero
eigenvalues do not affect the stability of a boundary equilibrium of
system \eqref{basicmodel}. This together with formula \eqref{detJ}
imply that an equilibrium of system \eqref{basicmodel} is stable if
and only if all the eigenvalues are non-positive.}
\end{Remark}

The following result will be used later in our study. The proof for
the discrete-time case may be found in \cite{Kingman}.

\begin{Proposition}\label{interiorstable}
Suppose $\Delta_1 \subset \Delta_2$ are two distinct invariant
subsets of $\Delta$ and the corresponding interior equilibria exist,
then they cannot be both stable.  In particular, if $\Delta$
contains a stable interior equilibrium, then no other equilibrium in
$\Delta$ can be stable.
\end{Proposition}

For example, for the four-allele case, $\Delta$ is a tetrahedron.
Let $\Delta_2$ be one of the four faces of $\Delta$ and suppose it
has a stable interior equilibrium. Then
Proposition~\ref{interiorstable} says that none of the vertices or
the edges of $\Delta_2$ (either can be $\Delta_1$ in
Proposition~\ref{interiorstable}) can contain a stable equilibrium.

\section{Three-allele Model}\label{3allele}
\setcounter{equation}{0} The invariant set $\Delta$ defined by
(\ref{deltadef}) is the triangle joining the vertices $(1,0), (0,1)$
and $(0,0)$. System (\ref{basicmodel}) is
\begin{equation}\label{3alleleeqn}
\dot{p}_i = p_i(r_i - r),\quad\quad i=1,2\,.
\end{equation}
We denote the three sides of $\Delta$ by $S_i = \{(p_1,p_2):p_i =
0\}, i=1,2,3$ and the boundary equilibrium on $S_i$ by $P_i$ if it
exists. Each side is a two-allele model so according to
Proposition~\ref{2alleleprop}, $P_i$ exists if $S_i$ is in the superior
case or inferior case and $P_i$ is stable and unstable,
respectively.  We now introduce some terminologies. Consider alleles
$A_i$ and $A_j$. If $r_{ij} = r_{ii}$, we say that $A_i$ is
completely dominant to $A_j$ and $A_j$ is recessive to $A_i$.
Partial dominance means $r_{ij}$ lies between $r_{ii}$ and $r_{jj}$
but not equal to their average and no dominance means $r_{ij} =
(r_{ii}+r_{jj})/2$.  For convenience, we say that $S_1$ is
completely dominant if $r_{23}=r_{22}$ or $r_{23}=r_{33}$ holds.
Likewise for edges $S_2$ and $S_3$.

\subsection{Non-Complete Dominance Case}
\label{noncompletedominance} In this subsection we assume that
$r_{ij}\neq r_{ii}, r_{jj}$. There are altogether $14$ patterns.
Five of them have one stable equilibrium, eight have two stable
equilibria, and one has three stable equilibria. It is difficult to
prove these results directly so we state four lemmas according to
how many $P_i$'s exist on the boundary of $\Delta$ and whether they
are in the intermediate, superior or inferior cases. The
classifications of the $14$ patterns will then follow easily from
these lemmas and are summarized in Tables \ref{tab1} -- \ref{tab3}.
For simplicity we assume that all the eigenvalues are nonzero (see
Remark \ref{remark3}). The proofs of the four lemmas are rather
technical and are given at the end of this subsection. Readers not
interested in proof may simply look at the tables.

\begin{Lemma}\label{noneexist}
Suppose none of the $P_i$'s exists. Then the interior equilibrium
does not exist.
\end{Lemma}

\begin{Lemma}\label{onePi}
Suppose only one $P_i$ exists. Then there are $8$ cases:
\begin{enumerate}
\item\,\,If $P_2$ exists but not $P_1, P_3$, then there are two cases:
(i) $S_2$ is in the inferior case, $P_2$ has two positive
eigenvalues and interior equilibrium does not exist, (ii) $S_2$ is
in the superior case, $P_2$ has two negative eigenvalues and
interior equilibrium does not exist.

\item\,\,If $P_1$ exists but not $P_2, P_3$, then
there are three cases: (i) $S_1$ is in the inferior case, $P_1$ has
two positive eigenvalues and interior equilibrium does not exist,
(ii) $S_1$ is in the superior case, $P_1$ has two negative
eigenvalues and interior equilibrium exists with one positive and
one negative eigenvalues, (iii) $S_1$ is in the superior case, $P_1$
has one negative and one positive eigenvalues and interior
equilibrium does not exist.

\item\,\,If $P_3$ exists but not $P_1, P_2$, then
there are three cases: (i) $S_3$ is in the superior case, $P_3$ has
two negative eigenvalues and interior equilibrium does not exist,
(ii) $S_3$ is in the inferior case, $P_3$ has two positive
eigenvalues and interior equilibrium exists with one positive and
one negative eigenvalue, (iii) $S_3$ is in the inferior case, $P_3$
has one positive and one negative eigenvalues and interior
equilibrium does not exist.
\end{enumerate}
\end{Lemma}

\begin{Lemma}\label{twoPi}
Suppose only two $P_i$'s exist.  Then there are $34$ cases:\\

\noindent A.\quad Suppose $P_2, P_3$ exist but not $P_1$. Then there
are $11$ cases:
\begin{enumerate}
\item\,\,If $S_2, S_3$ are both in the superior case, then there are $4$ cases.
\begin{enumerate}
\item\,\,$P_2$ or $P_3$ is stable with two negative eigenvalues and the other is unstable with one positive
and one negative eigenvalues. Interior equilibrium does not exist.
\item\,\,Both $P_2$ and $P_3$ are unstable with one positive and one negative eigenvalues.
Interior equilibrium exists and is stable.
\item\,\,Both $P_2$ and $P_3$ are stable with two negative eigenvalues. Interior equilibrium exists
with one positive and one negative eigenvalue.
\end{enumerate}

\item\,\,If $S_2, S_3$ are both in the inferior case, then there are $4$ cases.
\begin{enumerate}
\item\,\,Both $P_2$ and $P_3$ are unstable, one has two positive
eigenvalues and the other has one positive and one negative
eigenvalues. Interior equilibrium does not exist.
\item\,\,Both $P_2$ and $P_3$ are
unstable with two positive eigenvalues. Interior equilibrium exists
with one positive and one negative eigenvalues.
\item\,\,Both $P_2$ and $P_3$ are unstable with one positive and one negative eigenvalues. Interior
equilibrium exists with two positive eigenvalues.
\end{enumerate}

\item\,\,If $S_2$ is in the inferior case and $S_3$
is in the superior case, then there is only $1$ case. $P_2$ is
unstable with two positive eigenvalues and $P_3$ is stable with two
negative eigenvalues. Interior equilibrium does not exist.

\item\,\,If $S_2$ is in the superior case and $S_3$ is in the inferior
case, then there are $2$ cases.
\begin{enumerate}
\item\,\,$P_2$ is stable with two negative eigenvalues and $P_3$ is unstable one positive and one
negative eigenvalues. Interior equilibrium does not exist.
\item\,\,$P_2$ is stable with two negative eigenvalues, $P_3$ is unstable with two positive eigenvalues. Interior
equilibrium exists with one positive and one negative eigenvalue.
\end{enumerate}
\end{enumerate}
\smallskip

\noindent B.\quad Suppose $P_1, P_3$ exist but not $P_2$. Then there
are $12$ cases. The results are identical to Part A above except
$P_2$ and $P_1$ are interchanged in the statements of Part A. In
addition, there is an extra case in Part A 4 which we call 4(c):
$P_1$ is unstable with one positive and one negative eigenvalues,
$P_3$ is unstable with two positive eigenvalues. Interior
equilibrium does
not exist.\\

\noindent C.\quad Suppose $P_1, P_2$ exist but not $P_3$. Then there
are $11$ cases. The results are identical to Part A above except
$P_1$ is changed to $P_3$, $P_2$ is changed to $P_1$, and $P_3$ is
changed to $P_2$ in the statements of Part A. In addition, statement
4(a) in Part A is changed to the following: $P_1$ is unstable with
one positive and one negative eigenvalues, $P_2$ is unstable with
two positive eigenvalues. Interior equilibrium does not exist.
\end{Lemma}

\begin{Lemma} \label{allexist}
Suppose $P_1, P_2, P_3$ all exist. Then there are $26$ cases:
\begin{enumerate}
\item\,\,If $S_1, S_2, S_3$ are all in the inferior case, then there are $4$
cases:
\begin{enumerate}
\item\,\,$P_1, P_2, P_3$ are all unstable and each has one positive and one negative
eigenvalues. Interior equilibrium exists and has two positive
eigenvalues.
\item\,\,$P_1, P_2, P_3$ are all unstable. One of the $P_i$'s has two positive
eigenvalues and the rest have one positive and one negative
eigenvalues.  Interior equilibrium does not exist.
\end{enumerate}

\item\,\,If $S_i$ is in the superior case and the other two sides are
in the inferior case, then there are $9$ cases.
\begin{enumerate}
\item\,\,$P_i$ is stable with two negative eigenvalues and the other two
equilibria are unstable with two positive eigenvalues.  Interior
equilibrium exists with one positive and one negative eigenvalues.
\item\,\,$P_i$ is stable with two negative eigenvalues and the other two equilibria
are unstable, one has two positive eigenvalues while the other has
one positive and one negative eigenvalues. Interior equilibrium does
not exist.
\end{enumerate}

\item\,\,If $S_i$ is in the inferior case and the other two
sides are in the superior case, then there are $9$ cases.
\begin{enumerate}
\item\,\,$P_i$ is unstable with two positive eigenvalues and the other two
equilibria are stable with two negative eigenvalues. Interior
equilibrium exists with one positive and one negative eigenvalues.
\item\,\,$P_i$ is unstable with two positive eigenvalues, one of the other
equilibria is unstable with one positive and one negative
eigenvalues, and the third equilibrium is stable with two negative
eigenvalues. Interior equilibrium does not exist.
\end{enumerate}

\item\,\,If all three edges are in the superior case, then there are $4$
cases.
\begin{enumerate}
\item\,\,$P_1, P_2, P_3$ are all unstable with one positive and one negative
eigenvalues.  Interior equilibrium exists with two negative
eigenvalues.
\item\,\,One of the boundary equilibria is stable with two negative eigenvalues
and the other two are unstable with one positive and one negative
eigenvalues. Interior equilibrium does not exist.
\end{enumerate}
\end{enumerate}
\end{Lemma}

To prove Lemmas~\ref{noneexist}-\ref{allexist}, we need to work with
the differences of the $r_{ij}$'s. Let

\begin{equation}\label{greekdef}
\begin{array}{rcl}
\alpha_1 &=& r_{21} - r_{11}\quad \quad \alpha_2 = r_{31} -
r_{21}\quad \quad
\alpha_3 = r_{31} - r_{11} = \alpha_1 + \alpha_2\\
\beta_1 &=& r_{22} - r_{12}\quad \quad \beta_2 = r_{32} -
r_{22}\quad \quad
\beta_3 = r_{32} - r_{21} = \beta_1 + \beta_2\\
\gamma_1 &=& r_{23} - r_{13}\quad \quad \gamma_2 = r_{33} -
r_{23}\quad \quad \gamma_3 = r_{33} - r_{13} = \gamma_1 +
\gamma_2\,.
\end{array}
\end{equation}
\smallskip

From Proposition~\ref{interiorformula}, the interior equilibrium is
given by $p_i^* = U_i/\sum U_j, i=1,2,3$ where
\begin{eqnarray}
U_1 &=& \gamma_2 \beta_1-\gamma_1\beta_2 = \gamma_3 \beta_1 -
\gamma_1 \beta_3\,,\label{U1}\nonumber\\ U_2 &=& \alpha_2\gamma_1
-\alpha_1 \gamma_2    = \alpha_3 \gamma_1 -
\alpha_1\gamma_3\,,\label{U2} \label{Uformula}\\ U_3 &=&
\beta_2\alpha_1 -\beta_1 \alpha_2   = \beta_3 \alpha_1 - \beta_1
\alpha_3\,.\nonumber
\end{eqnarray}

Let $\Delta_{ij} = r_{ii}+r_{jj}-2r_{ij}$. The eigenvalues of $J$ at
the vertex and boundary equilibria are

\begin{center}
    \begin{tabular}{ | c | c | c | c |}
    \hline
    Equilibrium & Coordinates & First Eigenvalue & Second Eigenvalue\\ \hline
    Vertex 1& (0,0) & $-\gamma_3$ & $-\gamma_2$
    \\ \hline
    Vertex 2& (0,1) & $\beta_2$ & $-\beta_1$ \\ \hline
    Vertex 3 & (1,0) & $\alpha_1$ & $\alpha_3$ \\
    \hline
    $\quad P_1$ & $\left(0,\displaystyle\frac{\gamma_2}{\Delta_{23}}\right)$ & $\displaystyle\frac{-\beta_2\gamma_2}{\Delta_{23}}$ & $\displaystyle\frac{-U_1}{\Delta_{23}}$
    \\ \hline
    $\quad P_2$ & $\left(\displaystyle\frac{\gamma_3}{\Delta_{13}},0\right)$ & $\displaystyle\frac{-\alpha_3\gamma_3}{\Delta_{13}}$ & $\displaystyle\frac{-U_2}{\Delta_{13}}$\\ \hline
    $\quad P_3$ & $\left(\displaystyle\frac{\beta_1}{\Delta_{12}},\displaystyle\frac{-\alpha_1}{\Delta_{12}}\right)$ & $\displaystyle\frac{-\alpha_1\beta_1}{\Delta_{12}}$ & $\displaystyle\frac{-U_3}{\Delta_{12}}$ \\
    \hline
    \end{tabular}
\end{center}

The formulas for the eigenvalues of $J$ at the interior equilibrium
are too complex to be useful. However, we can determine their signs
from the determinant and trace of $J$ which are
\begin{eqnarray}\label{detJtraceJ}
\mbox{det}(J) &=& p_1^*p_2^*p_3^* (U_1+U_2+U_3) \,,\\
\mbox{tr}(J) &=& -\Delta_{13}(p_1^*)^2 - \Delta_{23}(p_2^*)^2 +
(\Delta_{12} - \Delta_{23} - \Delta_{13})p_1^*p_2^* +
\Delta_{13}p_1^* + \Delta_{23}p_2^*\nonumber\\
&=& \displaystyle\frac{U_1U_2\Delta_{12} + U_1 U_3\Delta_{13} +
U_2U_3\Delta_{23}}{(U_1+U_2+U_3)^2}\,. \nonumber
\end{eqnarray}
\smallskip

\noindent We only prove the simplest and the most difficult cases
since the rest are similar.\medskip

\noindent {\it Proof of Lemma~\ref{noneexist}.}\quad From
Proposition~\ref{interiorformula}, interior equilibrium exists if
and only if all the $U_i$'s are of the same sign.  Suppose all the
$U_i$'s are positive. Then from above, we have
\[
\gamma_1 \beta_2 < \gamma_2
\beta_1\,,\quad\quad\quad\;\gamma_2\alpha_1 < \alpha_2 \gamma_1\,,
\quad \;\;\;\;\beta_1 \alpha_2 < \beta_2 \alpha_1\,,
\]
\[
\gamma_1(-) < (-)(-)\,,\quad (-)(-) < \alpha_2\gamma_1 \,,\quad\quad
(-)\alpha_2 < (-)(-)\,.
\]
Therefore, $\alpha_2$ and $\gamma_1$ are of the same sign. They must
both be negative since $\alpha_2 + \gamma_1 = r_{23}-r_{12} =
\beta_1+\beta_2 < 0$. But then the above inequalities imply that
$\gamma_1/\gamma_2 < \beta_1/\beta_2,\, \alpha_1/\alpha_2 <
\gamma_1/\gamma_2,\, \beta_1/\beta_2 < \alpha_1/\alpha_2$ which is a
contradiction. Therefore, $U_i$'s cannot be all positive. The proof
of the fact that $U_i$'s cannot be all negative is similar. Hence,
interior equilibrium cannot exists. The proof of the lemma is
complete.\medskip

\noindent {\it Proof of Lemma~\ref{allexist}}\quad We only prove
parts 1 and 2 since the proofs of the rest are similar. Suppose all
three sides are in the inferior case so that $\Delta_{12}
,\Delta_{13} $ and $\Delta_{23}$ are all positive and the signs of
$\alpha_i, \beta_i, \gamma_i$ are
\[
\alpha_1 \;(-),\quad \alpha_2 \;(?),\quad \alpha_3\;(-),\quad
\beta_1 \;(+),\quad \beta_2 \;(-), \quad \beta_3\; (?),\quad
\gamma_1\; (?),\quad \gamma_2 \;(+),\quad \gamma_3\; (+)\,.
\]
From (\ref{Uformula}), we have
\begin{eqnarray*}
U_1 &=& (+)(+) - \gamma_1(-) = (+)(+) - \gamma_1 \beta_3\,,\\
U_2 &=& \alpha_2 \gamma_1 - (-)(+) = (-)\gamma_1 - (-)(+)\,,\\
U_3 &=& (-)(-) - (+)\alpha_2 = \beta_3(-) - (+)(-)\,.
\end{eqnarray*}
Since the signs of $\alpha_2, \beta_3$ and $\gamma_1$ are
indeterminate, we construct a table showing the signs of $U_i$'s
under all possible combinations of the signs of $\alpha_1, \beta_3$
and $\gamma_1$.
\begin{center}
\begin{tabular}{|c|l*{2}{c}|l*{3}{c}|c|}
\hline
Case & $\alpha_2$ & $\beta_3$ & $\gamma_1$ & $U_1$ & $U_2$  & $U_3$ &\\
\hline
 (i) & + & + & + & + & + & ? & \\
 (ii) & $-$ & + & + & + & ? & + & \\
 (iii) & $-$ & $-$ & + & + & ? & + & \\
 (iv) & $-$ & $-$ & $-$ & \,? & + & + & \\
(v) & + & + & $-$ & + & + & ? & \\
(vi) & + & $-$ & $-$ & \,? & + & + & \\
\hline
\end{tabular}
\end{center}
\smallskip

\noindent From the above table, the signs of the eigenvalues of $J$
at the boundary equilibria are
\[
P_1 \;\left\{\frac{+}{+},\frac{-U_1}{+}\right\},\quad\quad P_2
\;\left\{\frac{+}{+},\frac{-U_2}{+}\right\}, \quad \quad P_3
\;\left\{\frac{+}{+},\frac{-U_3}{+}\right\}\,.
\]

\noindent Recalling Lemma~\ref{allexist} Part 1 and referring to the
above table, we have\medskip

\noindent Cases (i) and (v).\hspace{.2in}If $U_{3}>0$, then Case
$(a)$ happens and if $U_{3}<0$, then Case $(b)$ happens.
\smallskip

\noindent Cases (ii) and (iii). \hspace{.09in}If $U_{2}>0$, then
Case $(a)$ happens and if $U_{2}<0$, then Case $(b)$ happens.
\smallskip

\noindent Cases iv) and (vi).\hspace{.09in} If $U_{1}>0$, then Case
$(a)$ happens and if $U_{1}<0$, then Case $(b)$ happens.\medskip

\noindent The proof of Lemma~\ref{allexist} Part 1 is complete. To
prove Part 2, suppose $i=1$ which means that $S_1$ is in the
superior case ($r_{33} < r_{22} < r_{23}$) and $S_2, S_3$ are in the
inferior case ($r_{13} < r_{33} < r_{11},\,r_{12} < r_{22} <
r_{11})$. Then $\Delta_{12},\, \Delta_{13}$ are positive and
$\Delta_{23}$ is negative. The signs of $\alpha_i, \beta_i,
\gamma_i$ are
\[
\alpha_1 \;(-),\quad \alpha_2 \;(?),\quad \alpha_3\;(-),\quad
\beta_1 \;(+),\quad \beta_2 \;(+), \quad \beta_3\; (+),\quad
\gamma_1\; (+),\quad \gamma_2 \;(-),\quad \gamma_3\; (+)\,.
\]
Note that $U_1 = \gamma_2\beta_1- \gamma_1\beta_2 = (-)(+)-(+)(+) <
0$ and the signs of the eigenvalues of $J$ at the boundary
equilibria are
\[
P_1 \;\left\{\frac{+}{-},\frac{+}{-}\right\},\quad\quad P_2
\;\left\{\frac{+}{+},\frac{-U_2}{+}\right\}, \quad\quad P_3
\;\left\{\frac{+}{+},\frac{-U_3}{+}\right\}\,.
\]
Therefore $P_1$ is always stable with two negative eigenvalues.
Suppose $\alpha_2>0$, then $U_3 = \beta_2 \alpha_1 - \beta_1
\alpha_2 < 0$ and $P_3 \;\{+,+\}$. If $U_{2}>0$, then $P_2
\;\{+,-\}$ and Case $(b)$ happens. If $U_{2}<0$, then $P_2
\;\{+,+\}$ and Case $(a)$ happens. Suppose $\alpha_2<0$, then $U_2 =
\alpha_2 \gamma_1 - \alpha_1 \gamma_2 <0$ and $P_2 \;\{+,+\}$. If
$U_{3}>0$, then $P_3 \;\{+,-\}$ and Case $(b)$ happens. If
$U_{3}<0$, then $P_3 \;\{+,+\}$ and Case $(a)$ happens. The proof of
Lemma~\ref{allexist} Part 2 is complete.\medskip

The $69$ cases in the above four lemmas are summarized in
Tables~\ref{tab1}-\ref{tab3} and the $14$ patterns are separated by
double lines in the tables.

\begin{table}[H]\label{oneequil}
\caption{One stable equilibrium: 30 cases 5 patterns. \,$+,-$ are
the signs of the eigenvalues of $J$ at the corresponding equilibrium
and x means equilibrium does not
exist.}\label{tab1}\vspace*{.15in}\centering
$\begin{array}{|l|cc|cc|cc|cc|cc|cc|cc|}

\hline $Lemma$ & \multicolumn{2}{|c|}{(0,0)} &
\multicolumn{2}{|c|}{(0,1)} & \multicolumn{2}{|c|}{(1,0)} &
\multicolumn{2}{|c|}{P_1} & \multicolumn{2}{|c|}{P_2} &
\multicolumn{2}{|c|}{P_3} & \multicolumn{2}{|c|}{$Interior$} \\
\hline

3.1 & + & + & - & + & - & - & $x$ & $x$ & $x$ & $x$ & $x$ & $x$ &
$x$ & $x$ \\ \hline

3.2.1(i) & - & + & - & + & - & - & $x$ & $x$ & + & + & $x$ & $x$ &
$x$ & $x$ \\ \hline

3.2.2(i) & + & - & - & + & - & - & + & + & $x$ & $x$ & $x$ & $x$ &
$x$ & $x$ \\ \hline

3.2.2(iii) & + & + & + & + & - & - & - & + & $x$ & $x$ & $x$ & $x$ &
$x$ & $x$ \\ \hline

3.3.B.4(c) & + & + & + & - & - & - & - & + & $x$ & $x$ & + & + & $x$
& $x$
\\ \hline

3.3.C.4(a) & - & + & + & + & - & - & - & + & + & + & $x$ & $x$ & $x$
& $x$
\\ \hline \hline

3.3.B.1(a) & + & + & + & + & + & - & - & - & $x$ & $x$ & - & + & $x$
& $x$
\\ \hline

3.3.C.1(a) & + & + & + & + & - & + & - & - & - & + & $x$ & $x$ & $x$
& $x$
\\ \hline

3.4.3(b) & + & + & + & - & - & + & - & - & - & + & + & + & $x$ & $x$
\\ \hline

3.4.3(b) & - & + & + & + & + & - & - & - & + & + & - & + & $x$ & $x$
\\ \hline

3.4.4(b) & + & + & + & + & + & + & - & - & - & + & - & + & $x$ & $x$
\\ \hline \hline

3.3.C.1(a) & + & + & + & + & - & + & - & + & - & - & $x$ & $x$ & $x$
& $x$
\\ \hline

3.3.C.3 & + & - & - & + & - & + & + & + & - & - & $x$ & $x$ & $x$ &
$x$
\\ \hline

3.4.3(b) & + & + & + & - & - & + & - & + & - & - & + & + & $x$ & $x$
\\ \hline

3.4.3(b) & + & - & - & + & + & + & + & + & - & - & - & + & $x$ & $x$
\\ \hline

3.4.4(b) & + & + & + & + & + & + & - & + & - & - & - & + & $x$ & $x$
\\ \hline

3.3.A.1(a) & + & + & - & + & + & + & $x$ & $x$ & - & - & - & + & $x$
& $x$
\\ \hline

3.2.1(ii) & + & + & - & + & - & + & $x$ & $x$ & - & - & $x$ & $x$ &
$x$ & $x$ \\ \hline \hline

3.2.3(i) & + & + & - & + & + & - & $x$ & $x$ & $x$ & $x$ & - & - &
$x$ & $x$ \\ \hline

3.3.A.1(a) & + & + & - & + & + & + & $x$ & $x$ & - & + & - & - & $x$
& $x$
\\ \hline

3.3.A.3 & - & + & - & + & + & - & $x$ & $x$ & + & + & - & - & $x$ &
$x$
\\ \hline

3.3.B.1(a) & + & + & + & + & + & - & - & + & $x$ & $x$ & - & - & $x$
& $x$
\\ \hline

3.3.B.3 & + & - & - & + & + & - & + & + & $x$ & $x$ & - & - & $x$ &
$x$
\\ \hline

3.4.3(b) & - & + & + & + & + & - & - & + & + & + & - & - & $x$ & $x$
\\ \hline

3.4.3(b) & + & - & - & + & + & + & + & + & - & + & - & - & $x$ & $x$
\\ \hline

3.4.4(b) & + & + & + & + & + & + & - & + & - & + & - & - & $x$ & $x$
\\ \hline \hline

3.3.A.1(b) & + & + & - & + & + & + & $x$ & $x$ & - & + & - & + & - & - \\
\hline

3.3.B.1(b) & + & + & + & + & + & - & - & + & $x$ & $x$ & - & + & - & - \\
\hline

3.3.C.1(b) & + & + & + & + & - & + & - & + & - & + & $x$ & $x$ & - & - \\
\hline

3.4.4(a) & + & + & + & + & + & + & - & + & - & + & - & + & - & - \\
\hline

\end{array}$
\smallskip
\end{table}

\begin{table}[H]
\caption{Two stable equilibria: 35 cases 8 patterns. \,$+,-$ are the
signs of the eigenvalues of $J$ at the corresponding equilibrium and
x means equilibrium does not exist.}\label{tab2}\vspace*{.15in}\centering
$\begin{array}{|l|cc|cc|cc|cc|cc|cc|cc|}

\hline $Lemma$ & \multicolumn{2}{|c|}{(0,0)} &
\multicolumn{2}{|c|}{(0,1)} & \multicolumn{2}{|c|}{(1,0)} &
\multicolumn{2}{|c|}{P_1} & \multicolumn{2}{|c|}{P_2} &
\multicolumn{2}{|c|}{P_3} & \multicolumn{2}{|c|}{$Interior$} \\
\hline

3.2.3(ii) & + & + & - & - & - & - & $x$ & $x$ & $x$ & $x$ & + & + &
+ & -
\\ \hline

3.2.3(iii) & + & + & - & - & - & - & $x$ & $x$ & $x$ & $x$ & + & - &
$x$ & $x$ \\ \hline

3.3.B.2(a) & + & - & - & - & - & - & + & - & $x$ & $x$ & + & + & $x$
& $x$
\\ \hline

3.3.B.2(a) & + & - & - & - & - & - & + & + & $x$ & $x$ & + & - & $x$
& $x$
\\ \hline

3.3.B.2(b) & + & - & - & - & - & - & + & + & $x$ & $x$ & + & + & + & - \\
\hline

3.3.B.2(c) & + & - & - & - & - & - & + & - & $x$ & $x$ & + & - & + & + \\
\hline

3.3.A.2(a) & - & + & - & - & - & - & $x$ & $x$ & + & - & + & + & $x$
& $x$
\\ \hline

3.3.A.2(a) & - & + & - & - & - & - & $x$ & $x$ & + & + & + & - & $x$
& $x$
\\ \hline

3.3.A.2(b) & - & + & - & - & - & - & $x$ & $x$ & + & + & + & + & + & - \\
\hline

3.3.A.2(c) & - & + & - & - & - & - & $x$ & $x$ & + & - & + & - & + & + \\
\hline \hline

3.3.C.2(a) & - & - & - & + & - & - & + & - & + & + & $x$ & $x$ & $x$
& $x$
\\ \hline

3.3.C.2(a) & - & - & - & + & - & - & + & + & + & - & $x$ & $x$ & $x$
& $x$
\\ \hline

3.3.C.2(b) & - & - & - & + & - & - & + & + & + & + & $x$ & $x$ & + & - \\
\hline

3.3.C.2(c) & - & - & - & + & - & - & + & - & + & - & $x$ & $x$ & + & + \\
\hline \hline

3.2.2(ii) & + & + & + & + & - & - & - & - & $x$ & $x$ & $x$ & $x$ &
+ & -
\\ \hline

3.3.B.4(a) & + & + & + & - & - & - & - & - & $x$ & $x$ & + & - & $x$
& $x$
\\ \hline

3.3.B.4(b) & + & + & + & - & - & - & - & - & $x$ & $x$ & + & + & + & - \\
\hline

3.4.2(a) & - & + & + & - & - & - & - & - & + & + & + & + & + & - \\
\hline

3.4.2(b) & - & + & + & - & - & - & - & - & + & + & + & - & $x$ & $x$
\\ \hline

3.4.2(b) & - & + & + & - & - & - & - & - & + & - & + & + & $x$ & $x$
\\ \hline

3.3.C.4(b) & - & + & + & + & - & - & - & - & + & + & $x$ & $x$ & + & - \\
\hline \hline

3.3.A.4(a) & + & + & - & - & - & + & $x$ & $x$ & - & - & + & - & $x$
& $x$
\\ \hline

3.3.A.4(b) & + & + & - & - & - & + & $x$ & $x$ & - & - & + & + & + & - \\
\hline

3.4.2(a) & + & - & - & - & - & + & + & + & - & - & + & + & + & - \\
\hline

3.4.2(b) & + & - & - & - & - & + & + & + & - & - & + & - & $x$ & $x$
\\ \hline

3.4.2(b) & + & - & - & - & - & + & + & - & - & - & + & + & $x$ & $x$
\\ \hline \hline

3.4.2(a) & - & - & - & + & + & - & + & + & + & + & - & - & + & - \\
\hline

3.4.2(b) & - & - & - & + & + & - & + & + & + & - & - & - & $x$ & $x$
\\ \hline

3.4.2(b) & - & - & - & + & + & - & + & - & + & + & - & - & $x$ & $x$
\\ \hline \hline

3.3.C.1(c) & + & + & + & + & - & + & - & - & - & - & $x$ & $x$ & + & - \\
\hline

3.4.3(a) & + & + & + & - & - & + & - & - & - & - & + & + & + & - \\
\hline \hline

3.3.B.1(c) & + & + & + & + & + & - & - & - & $x$ & $x$ & - & - & + & - \\
\hline

3.4.3(a) & - & + & + & + & + & - & - & - & + & + & - & - & + & - \\
\hline \hline

3.3.A.1(c) & + & + & - & + & + & + & $x$ & $x$ & - & - & - & - & + & - \\
\hline

3.4.3(a) & + & - & - & + & + & + & + & + & - & - & - & - & + & - \\
\hline

\end{array}$
\smallskip
\end{table}

\begin{table}[H]
\caption{Four cases 1 pattern with three stable equilibria. \,$+,-$
are the signs of the eigenvalues of $J$ at the corresponding
equilibrium and x means equilibrium does not
exist.}\label{tab3}\vspace*{.15in}\centering
$\begin{array}{|l|cc|cc|cc|cc|cc|cc|cc|}

\hline

$Lemma$ & \multicolumn{2}{|c|}{(0,0)} & \multicolumn{2}{|c|}{(0,1)}
& \multicolumn{2}{|c|}{(1,0)} & \multicolumn{2}{|c|}{P_1} &
\multicolumn{2}{|c|}{P_2} &
\multicolumn{2}{|c|}{P_3} & \multicolumn{2}{|c|}{$Interior$} \\
\hline

3.4.1(a) & - & - & - & - & - & - & + & - & + & - & + & - & + & + \\
\hline

3.4.1(b) & - & - & - & - & - & - & + & + & + & - & + & - & $x$ & $x$
\\ \hline

3.4.1(b) & - & - & - & - & - & - & + & - & + & + & + & - & $x$ & $x$
\\ \hline

3.4.1(b) & - & - & - & - & - & - & + & - & + & - & + & + & $x$ & $x$
\\ \hline

\end{array}$
\smallskip
\end{table}

Knowing the patterns is not enough to predict the asymptotic
behavior of solutions of \eqref{3alleleeqn}.  We also need to know
the separatrices for each pattern. Separatrices are curves which
divided $\Delta$ into separate regions each containing a stable
equilibrium. If the initial condition lies in one of these regions,
then the solutions of \eqref{3alleleeqn} will converge to the stable
equilibrium in that region as $t \rightarrow \infty$. The $14$
patterns together with their separatrices are shown in pictures in
the following theorem. In these pictures, big dots and small dots
represent  stable and unstable equilibria, respectively. Arrows
indicate the directions of flow of the solutions.\medskip

\begin{Theorem}\label{3allelemain}
For the three-allele case without complete dominance, $14$ patterns
may occur.
\begin{enumerate}
\item\,\,There are $5$ patterns with one stable equilibrium. They are
$\{(1,0)\}$, $\{P_1\}$, $\{P_2\}$, $\{P_3\}$, and the interior
equilibrium, respectively. There is no separatrix. (See Table \ref{tab1}.)

\item\,\,There are $8$ patterns with two stable equilibria. (See Table \ref{tab2}.) They
are\medskip

(a) $\{(0,1),(1,0)\}:$ The separatrix may behave in $10$ different
ways. See Fig. \ref{fig1} and \ref{fig2}.

\begin{figure}[H]
\centering
\begin{tabular}{ccccc}
{\includegraphics[height=1.0in]{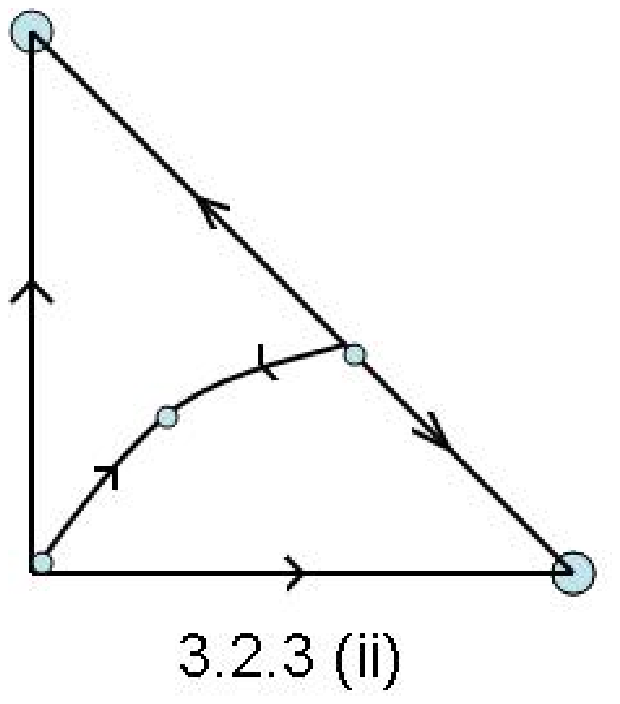}} &
{\includegraphics[height=1.0in]{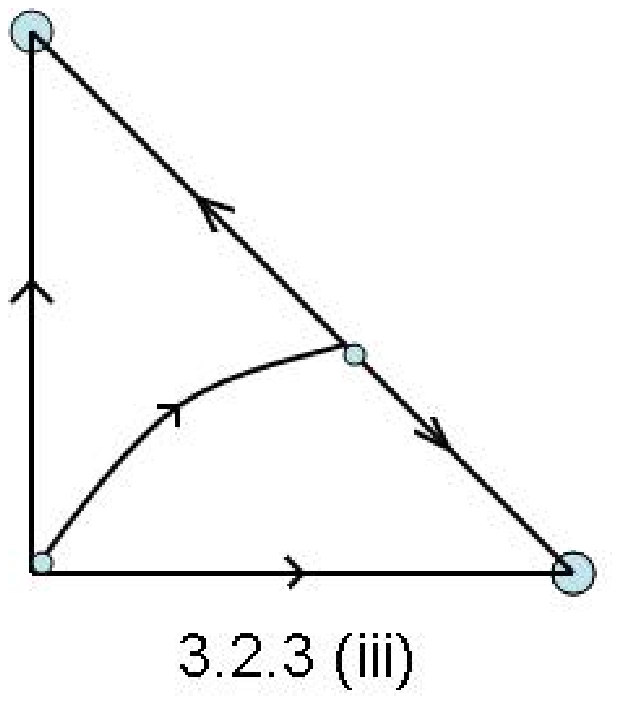}} &
{\includegraphics[height=1.0in]{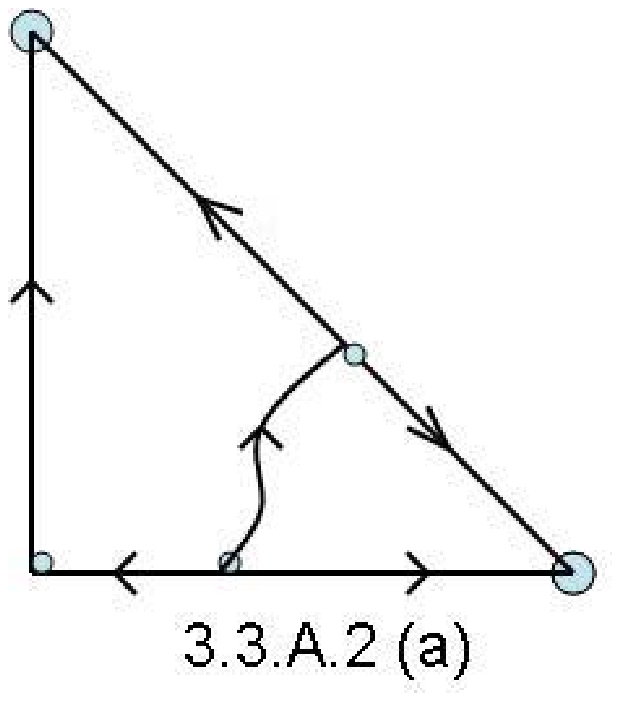}} &
{\includegraphics[height=1.0in]{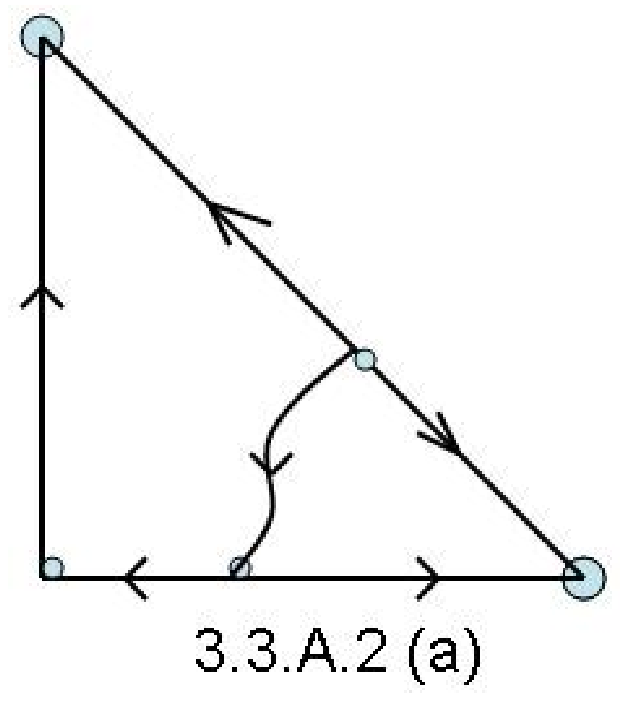}} &
{\includegraphics[height=1.0in]{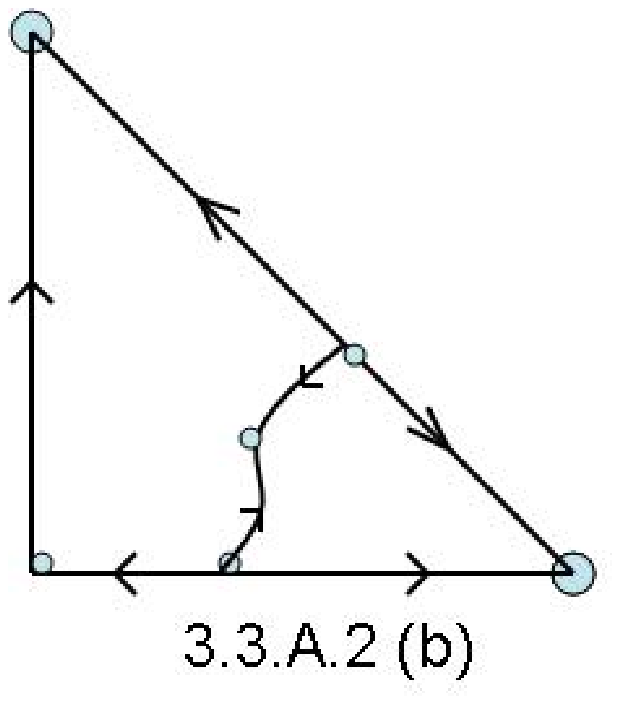}}
\end{tabular}
\caption{$\{(0,1),(1,0)\}$} \label{fig1}
\end{figure}

\begin{figure}[H]
\centering
\begin{tabular}{ccccc}
{\includegraphics[height=1.0in]{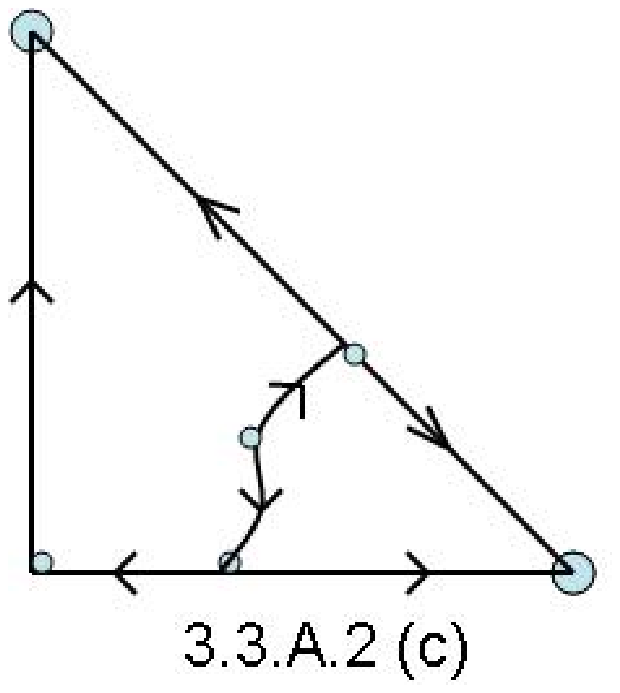}} &
{\includegraphics[height=1.0in]{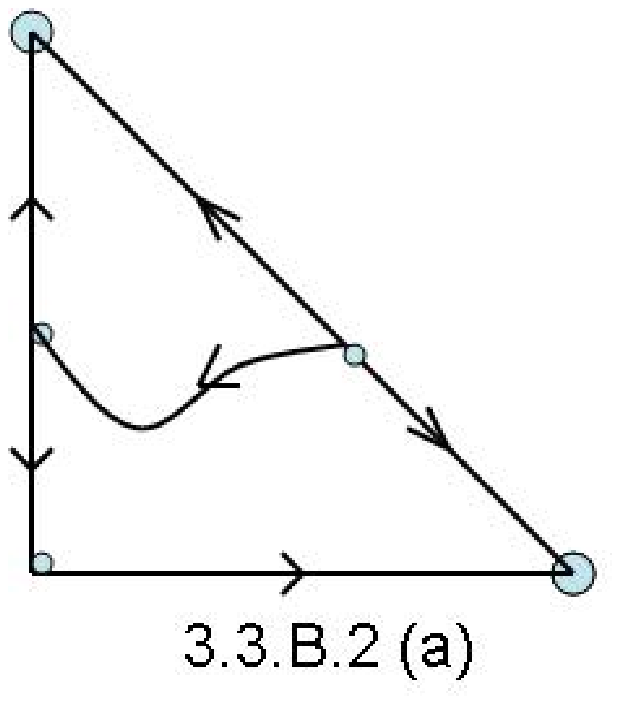}} &
{\includegraphics[height=1.0in]{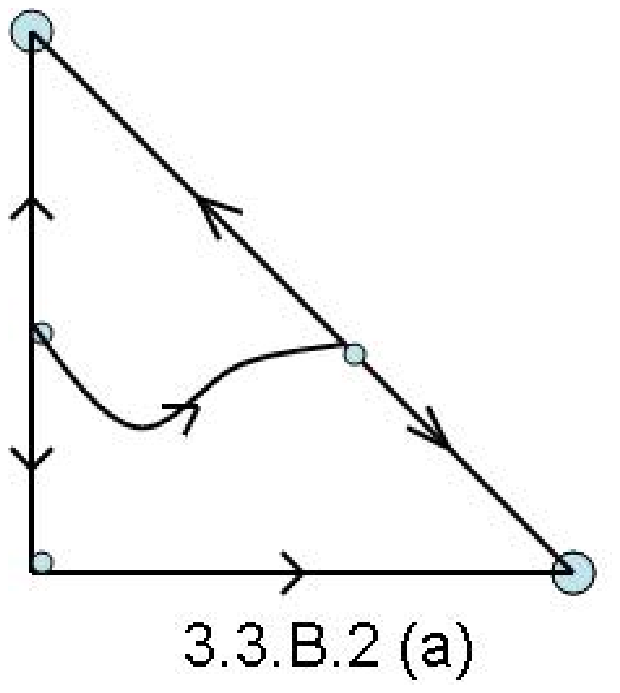}} &
{\includegraphics[height=1.0in]{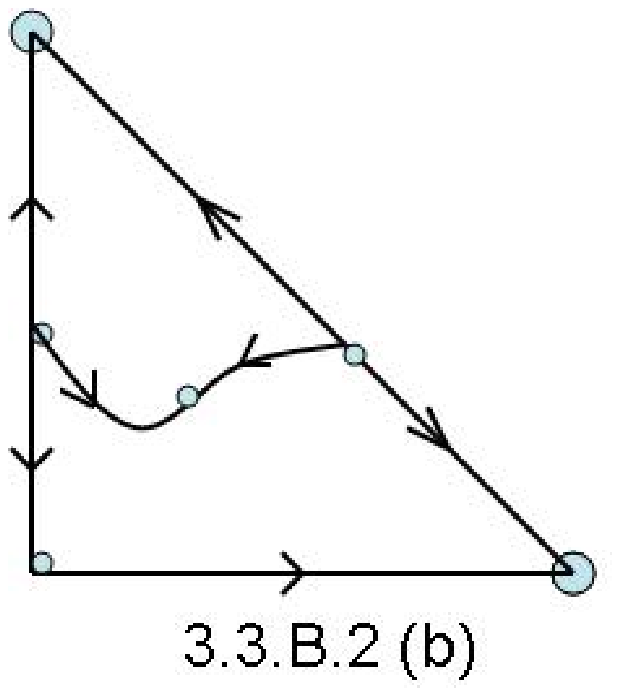}} &
{\includegraphics[height=1.0in]{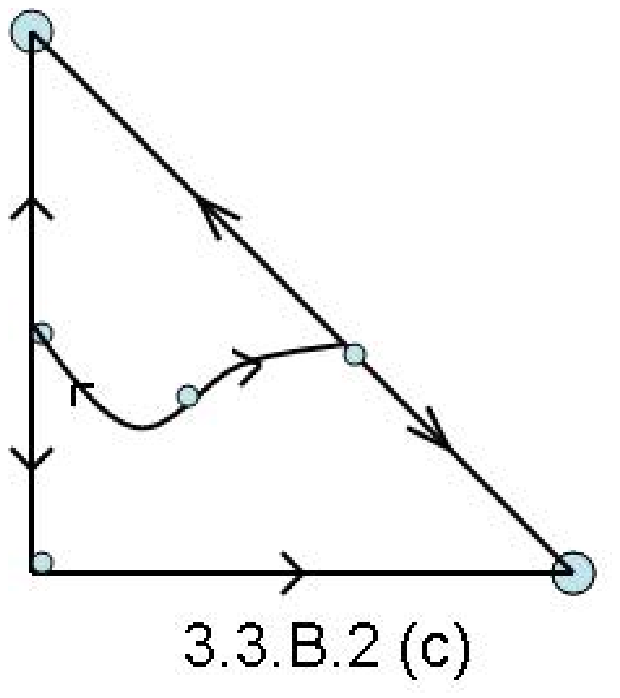}}
\end{tabular}
\caption{$\{(0,1),(1,0)\}$} \label{fig2}
\end{figure}


(b) $\{(0,0),(1,0)\}:$ The separatrix may behave in $4$ different
ways. See Fig. \ref{fig3}.

\begin{figure}[H]
\centering
\begin{tabular}{cccc}
{\includegraphics[height=1.0in]{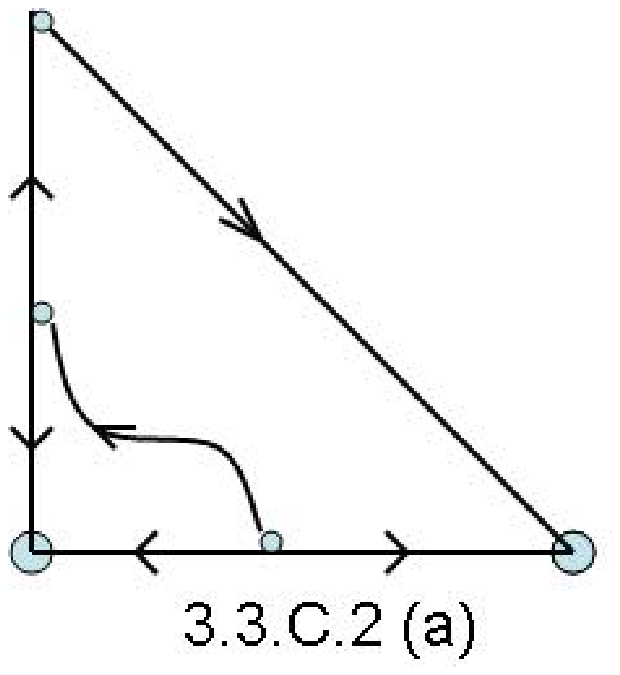}} &
{\includegraphics[height=1.0in]{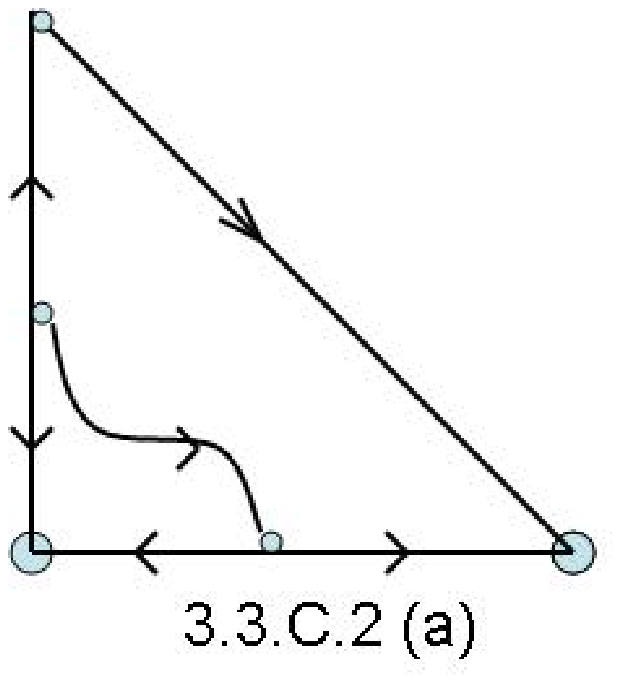}} & 
{\includegraphics[height=1.0in]{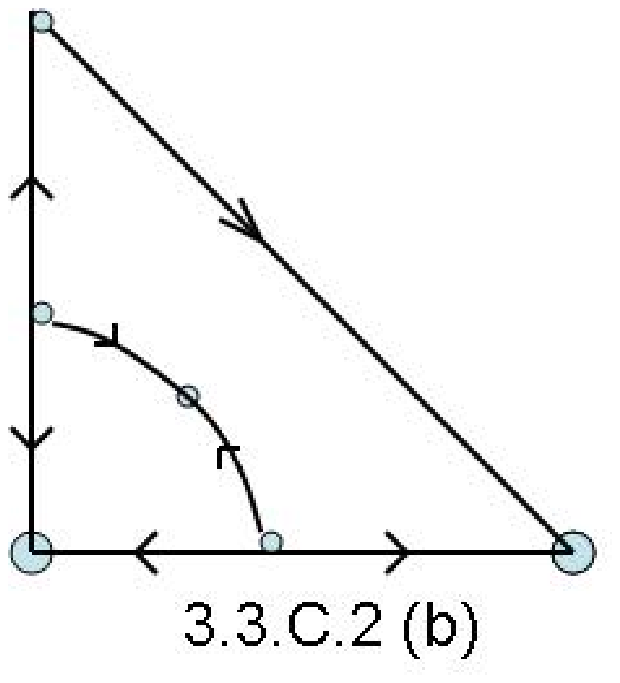}} &
{\includegraphics[height=1.0in]{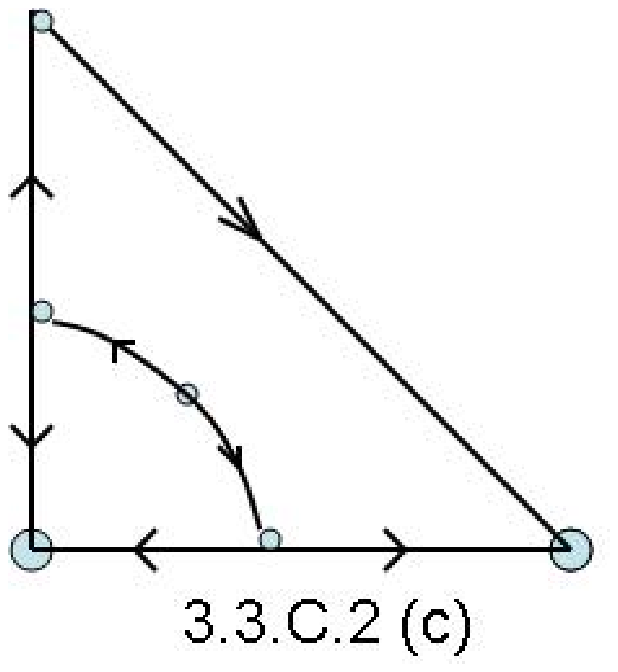}}
\end{tabular}
\caption{$\{(0,0),(1,0)\}$} \label{fig3}
\end{figure}

(c) $\{(1,0),P_1\}:$ The separatrix may behave in $7$ different
ways. See Fig. \ref{fig4} and \ref{fig5}.

\begin{figure}[H]
\centering
\begin{tabular}{cccc}
{\includegraphics[height=1.0in]{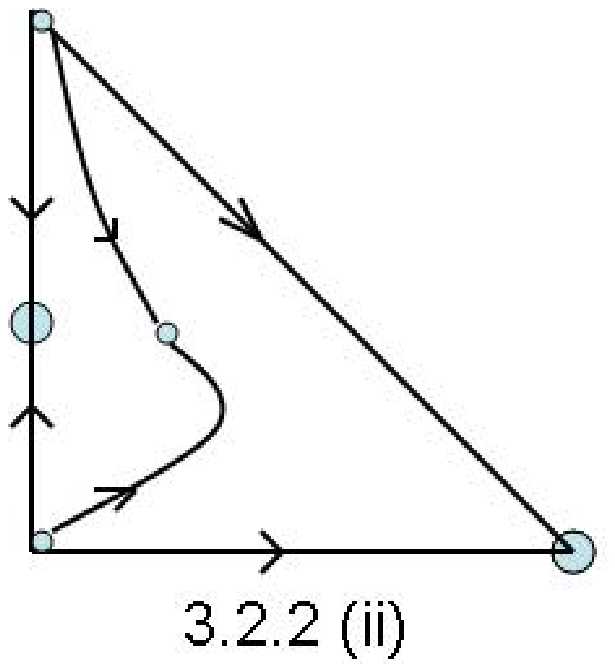}} &
{\includegraphics[height=1.0in]{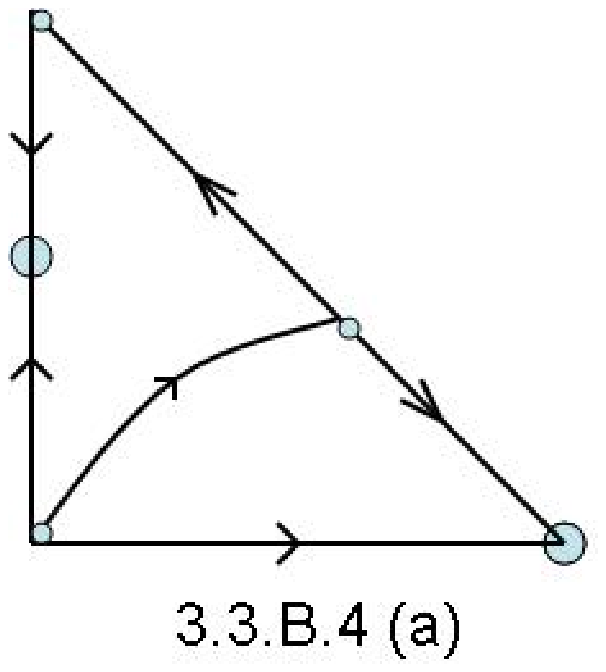}} &
{\includegraphics[height=1.0in]{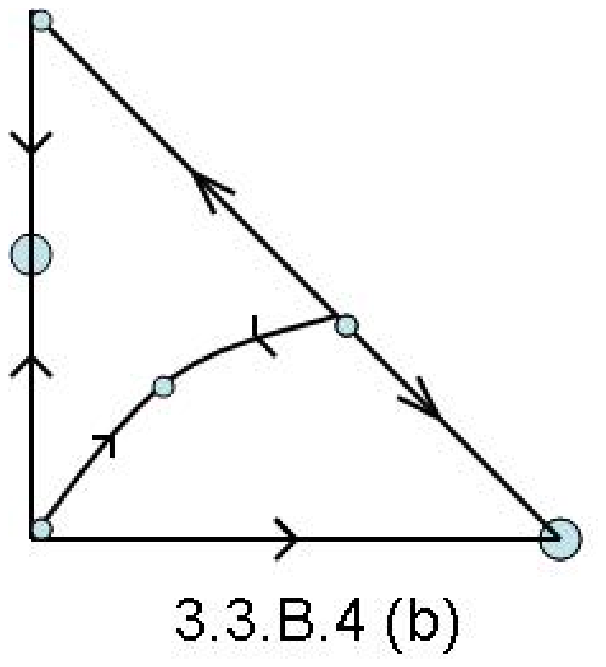}} &
{\includegraphics[height=1.0in]{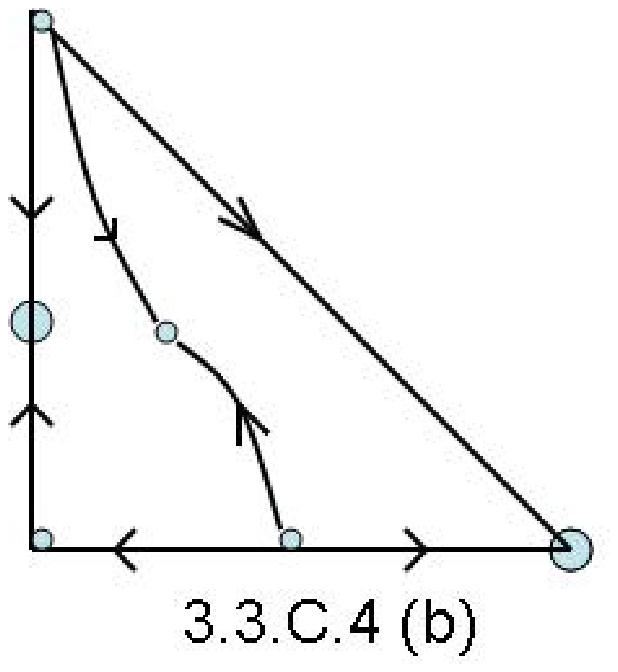}}
\end{tabular}
\caption{$\{(1,0),P_1\}$} \label{fig4}
\end{figure}

\begin{figure}[H]
\centering
\begin{tabular}{ccc}
{\includegraphics[height=1.0in]{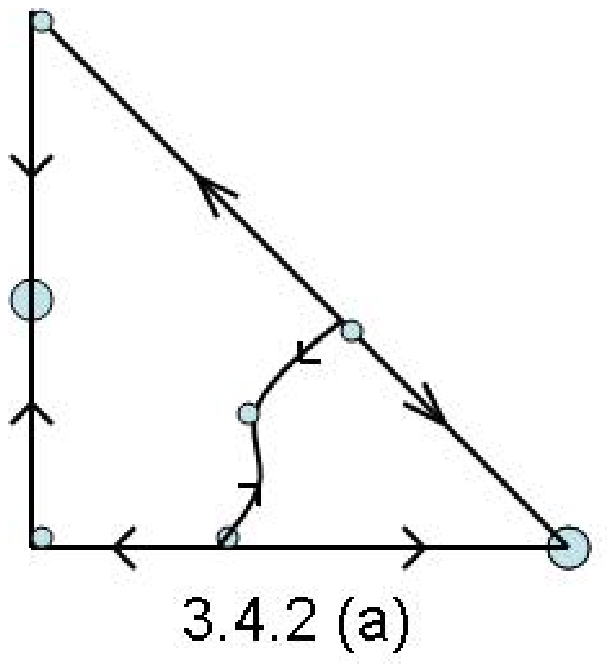}} &
{\includegraphics[height=1.0in]{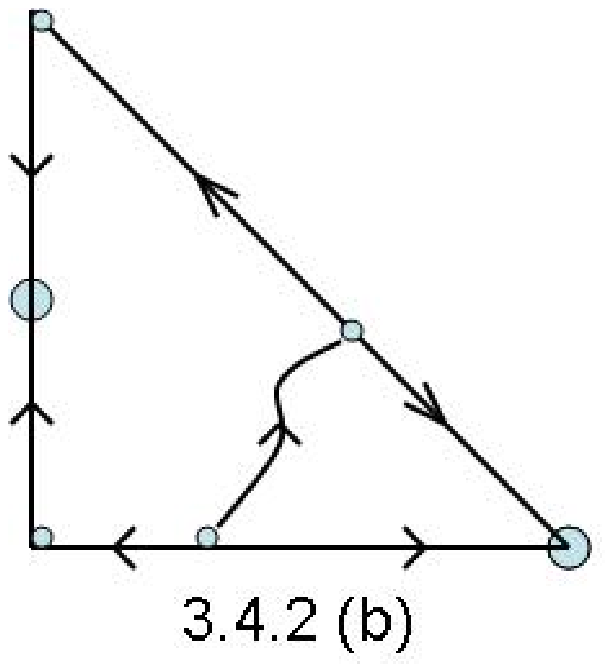}} &
{\includegraphics[height=1.0in]{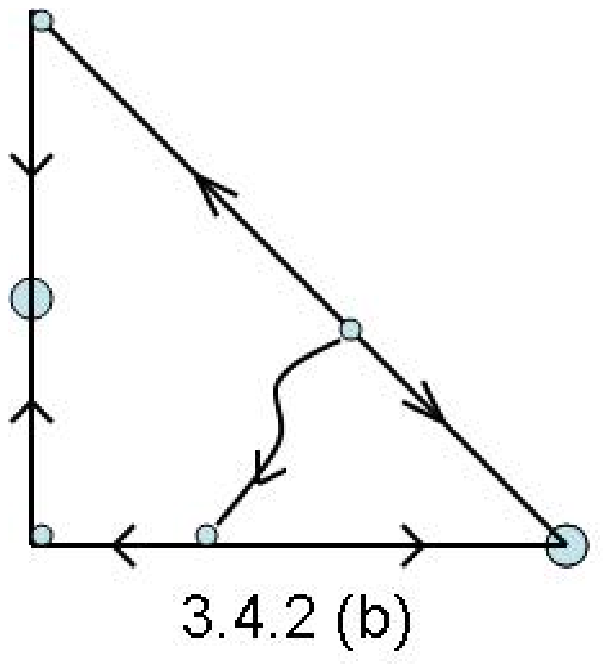}}
\end{tabular}
\caption{$\{(1,0),P_1\}$} \label{fig5}
\end{figure}

(d) $\{(0,1),P_2\}:$ The separatrix may behave in $5$ different
ways. See Fig. \ref{fig6}.

\begin{figure}[H]
\centering
\begin{tabular}{ccccc}
{\includegraphics[height=1.0in]{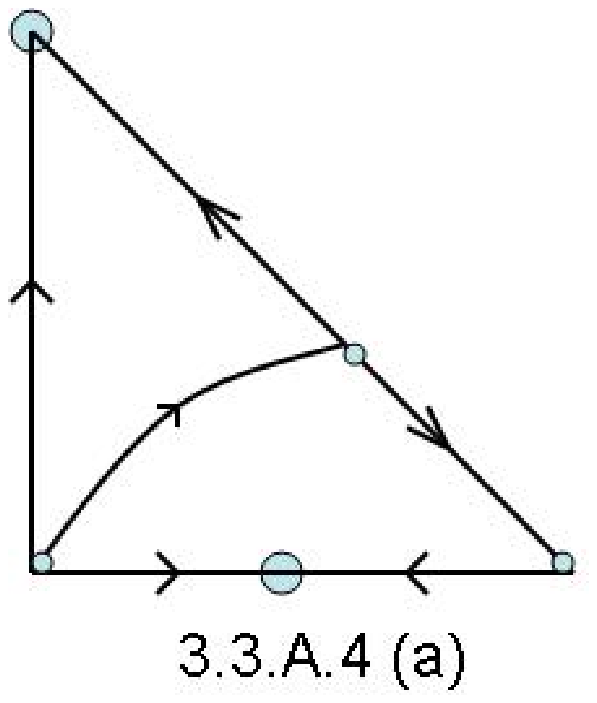}} &
{\includegraphics[height=1.0in]{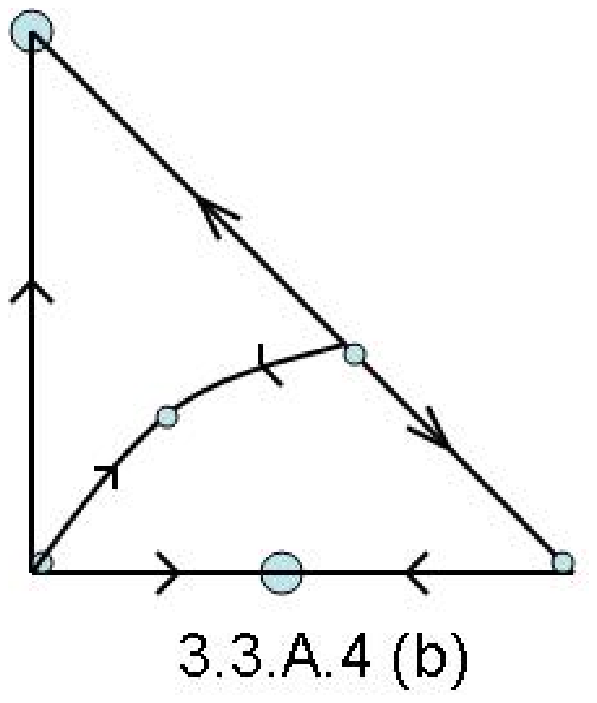}} &
{\includegraphics[height=1.0in]{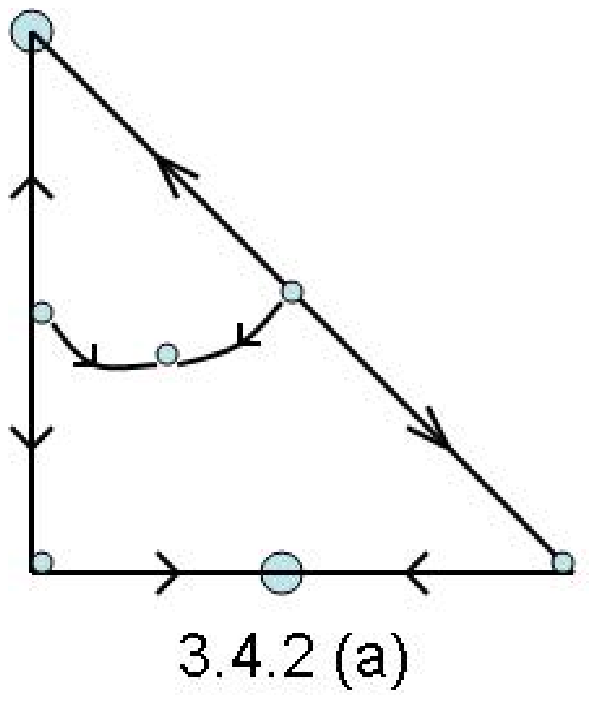}} &
{\includegraphics[height=1.0in]{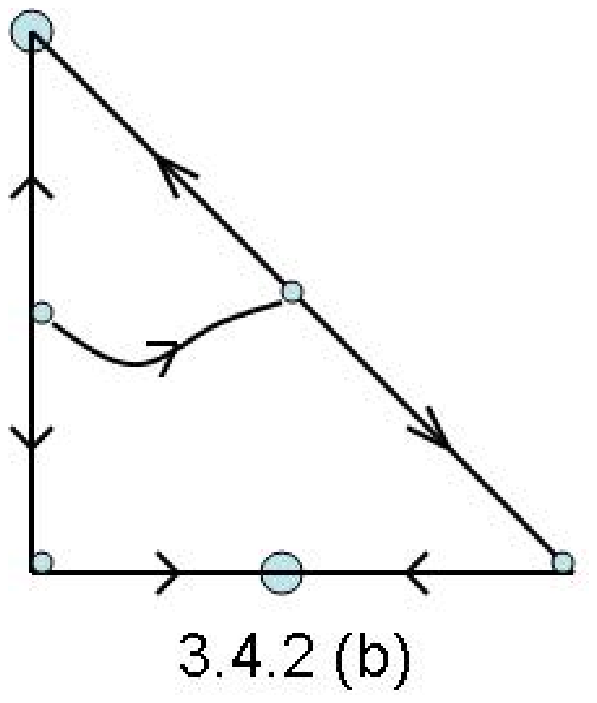}} &
{\includegraphics[height=1.0in]{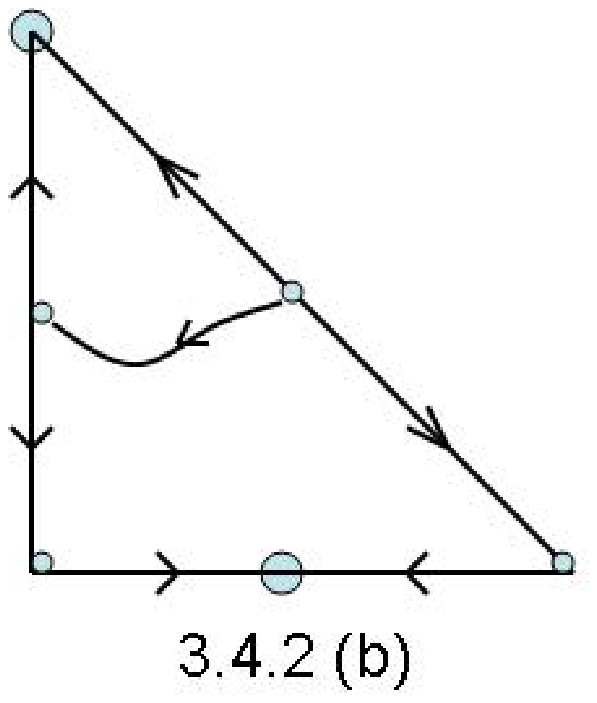}}
\end{tabular}
\caption{$\{(0,1),P_2\}$} \label{fig6}
\end{figure}
 \newpage

(e) $\{(0,0),P_3\}:$ The separatrix may behave in $3$ different
ways. See Fig. \ref{fig7}.

\begin{figure}[H]
\centering
\begin{tabular}{ccc}
{\includegraphics[height=1.0in]{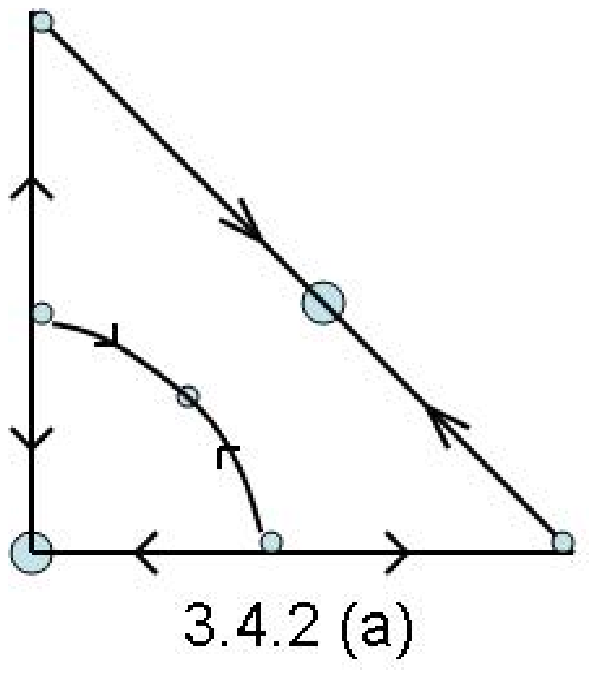}} &
{\includegraphics[height=1.0in]{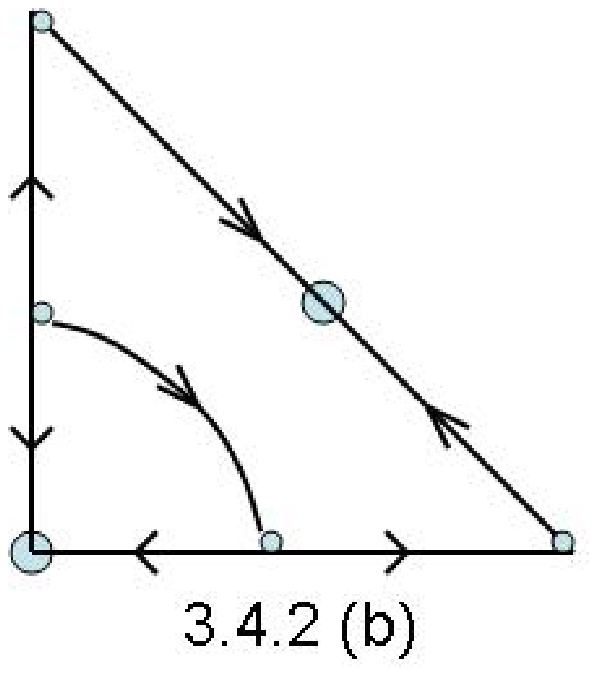}} &
{\includegraphics[height=1.0in]{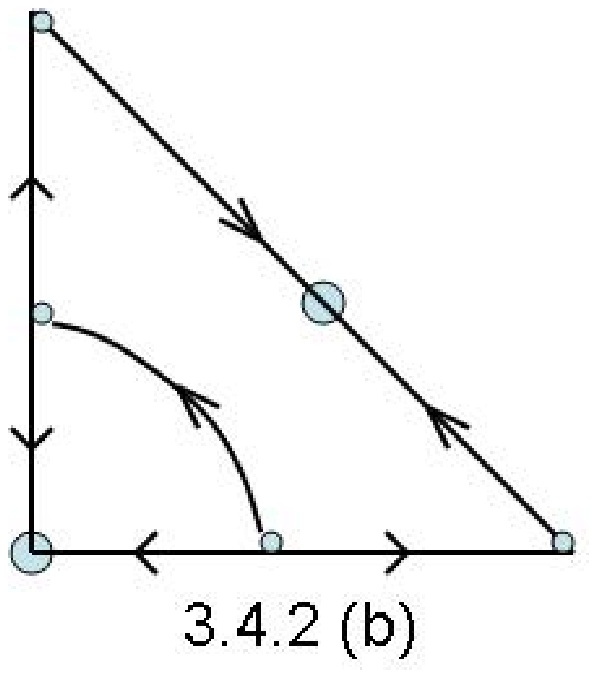}}
\end{tabular}
\caption{$\{(0,0),P_3\}$} \label{fig7}
\end{figure}

(f) $\{P_1,P_2\}:$ The separatrix may behave in $2$ different ways.
See Fig. \ref{fig8}.

\begin{figure}[H]
\centering
\begin{tabular}{cc}
{\includegraphics[height=1.0in]{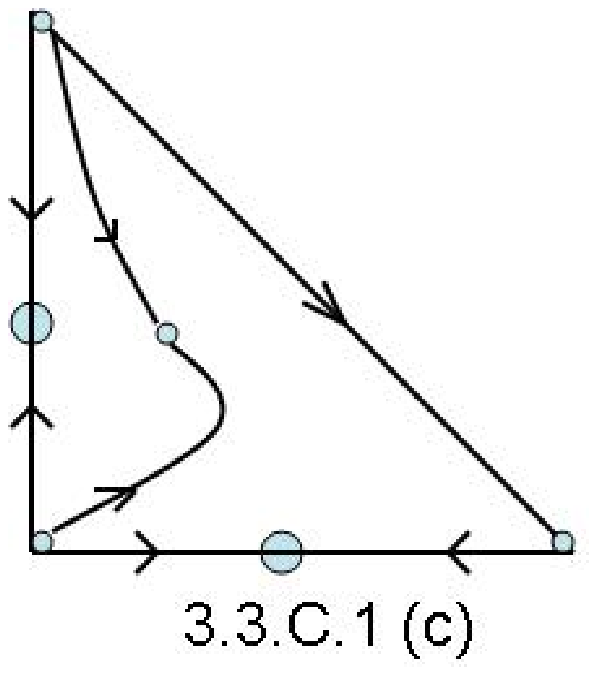}} &
{\includegraphics[height=1.0in]{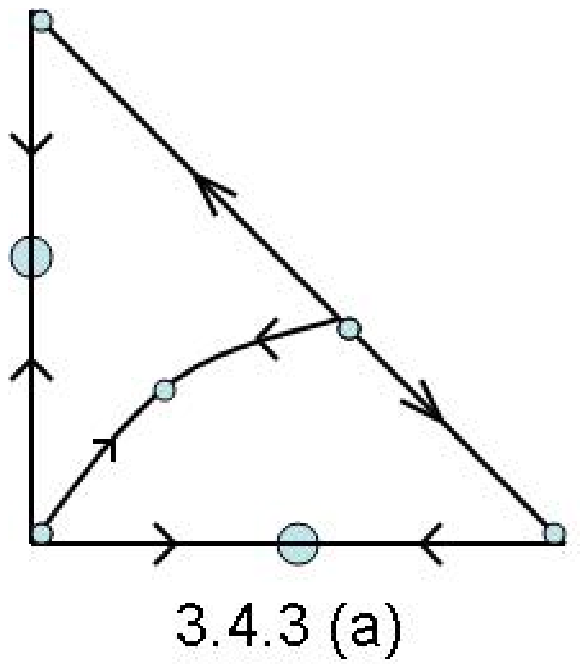}}
\end{tabular}
\caption{$\{P_1,P_2\}$} \label{fig8}
\end{figure}

(g) $\{P_1,P_3\}:$ The separatrix may behave in $2$ different ways.
See Fig. \ref{fig9}.

\begin{figure}[H]
\centering
\begin{tabular}{cc}
{\includegraphics[height=1.0in]{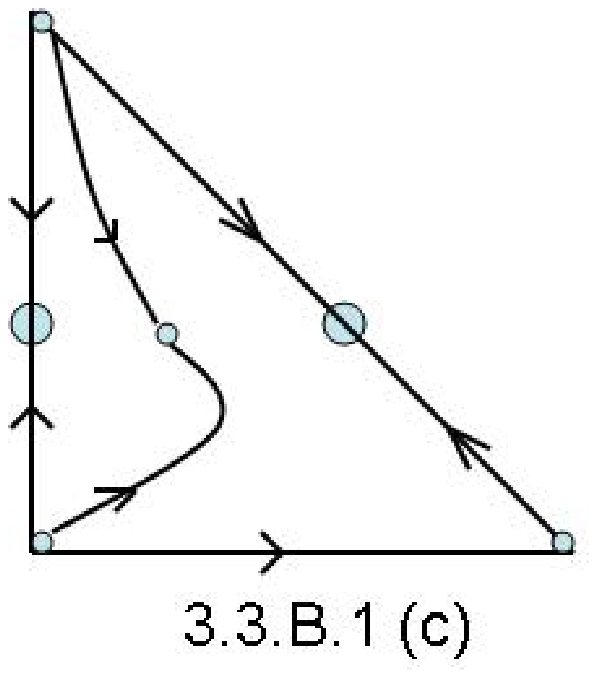}} &
{\includegraphics[height=1.0in]{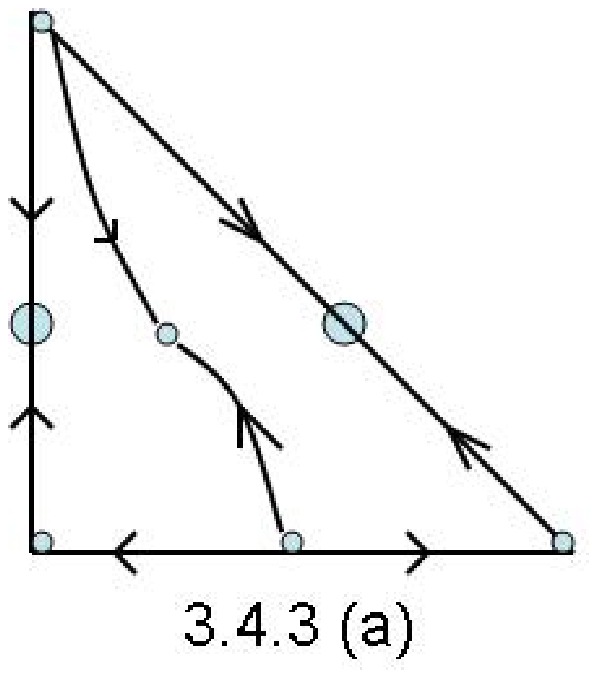}}
\end{tabular}
\caption{$\{P_1,P_3\}$} \label{fig9}
\end{figure}

(h) $\{P_2,P_3\}:$ The separatrix may behave in $2$ different ways.
See Fig. \ref{fig10}.

\begin{figure}[H]
\centering
\begin{tabular}{cc}
{\includegraphics[height=1.0in]{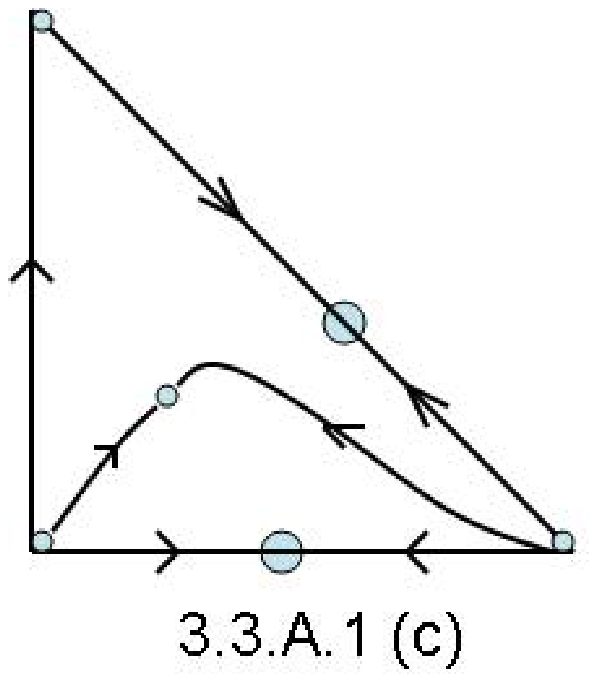}} &
{\includegraphics[height=1.0in]{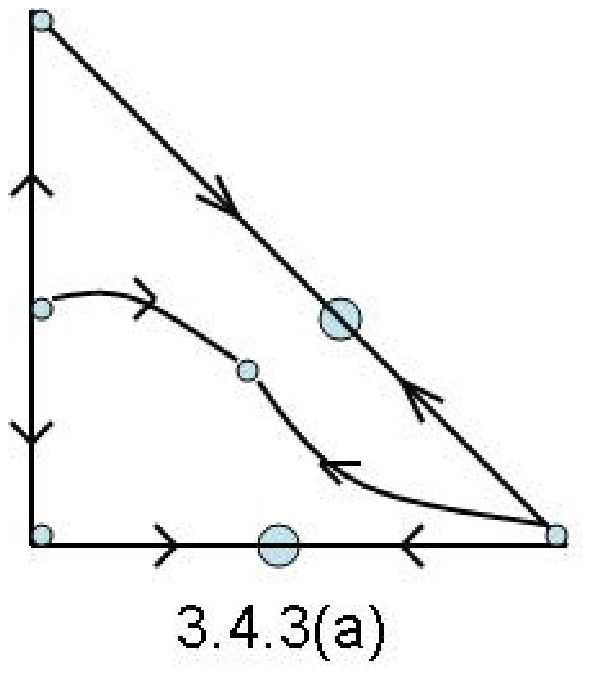}}
\end{tabular}
\caption{$\{P_2,P_3\}$} \label{fig10}
\end{figure}

\item\,\,There is only one pattern with three stable equilibria which must be the three
vertices. (See Table \ref{tab3}.) The separatrix may behave in $4$ different
ways. See Fig. \ref{fig11}.

\begin{figure}[H]
\centering
\begin{tabular}{cccc}
{\includegraphics[height=1.0in]{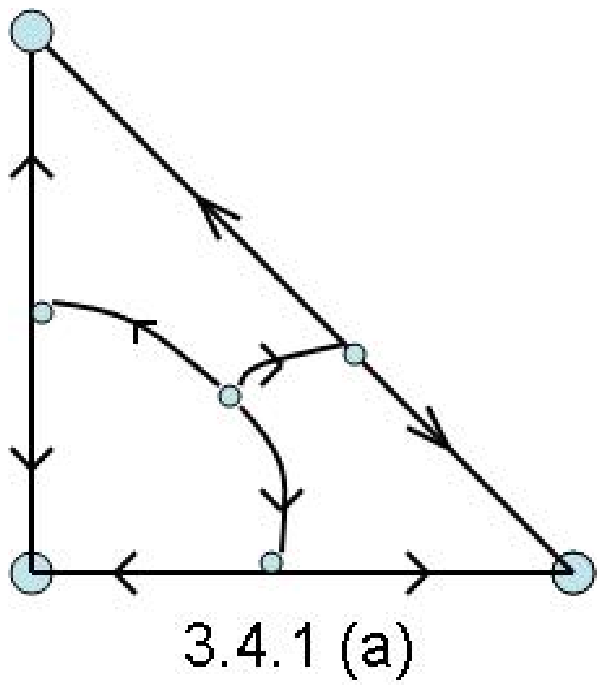}} &
{\includegraphics[height=1.0in]{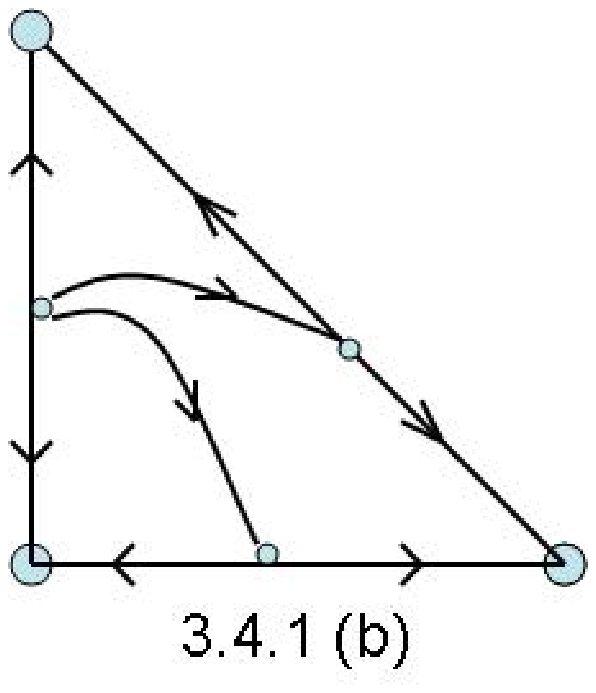}} &
{\includegraphics[height=1.0in]{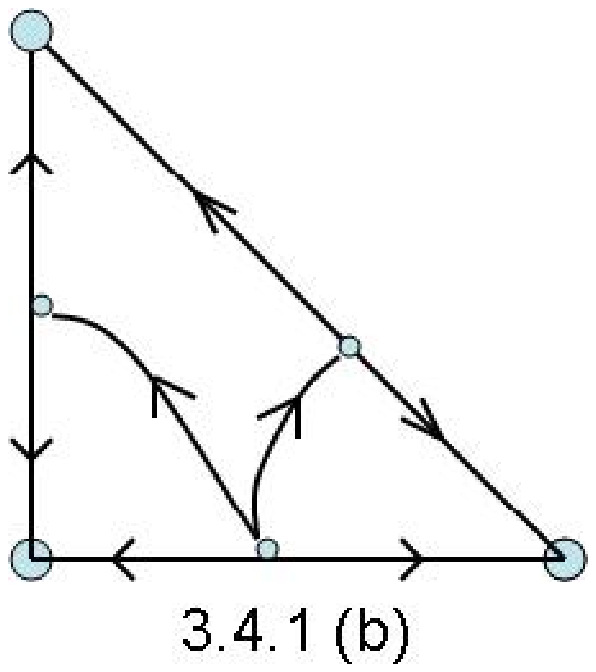}} &
{\includegraphics[height=1.0in]{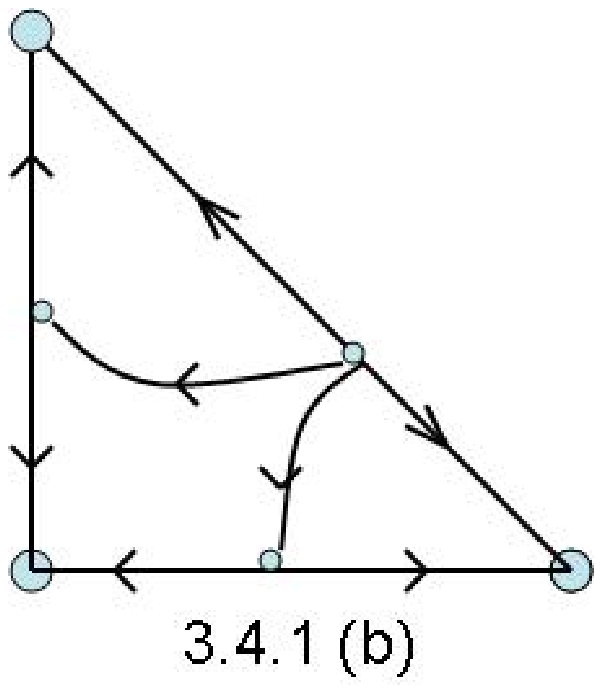}}
\end{tabular}
\caption{$\{(1,0),(0,1),(0,0)\}$} \label{fig11}
\end{figure}
\end{enumerate}
\end{Theorem}

\subsection{Complete Dominance Case}
In this subsection, we consider the case of complete dominance on at
least one side of $\Delta$. There are altogether $26$ cases to
consider and the number of patterns is greatly reduced in each
case.\medskip

\noindent (i)\,\,Complete dominance on one side.\smallskip

There are $6$ ways to assume complete dominance on one
side.\smallskip

(a) Suppose $S_3$ is completely dominant ($r_{12} = r_{11}$ or
$r_{12} = r_{22}$).\,\,The dynamics of \eqref{3alleleeqn} on $S_3$
is the same as $S_3$ being in the intermediate case. The dynamics on
the other two edges will not be affected if $S_3$ is changed from
intermediate case to complete dominance case. Therefore, to find all
the patterns, we go through every case of $S_3$ being in the
intermediate case in Tables 1-3 and examine how the signs of
$U_{i}'s$ will change when the condition $r_{22}<r_{12}<r_{11}$ is
replaced with $r_{12} = r_{11}$ or $r_{12}= r_{22}$.
There are $17$ such cases which generate $7$ patterns:
\[\{(1,0)\},\quad\{P_1\},\quad\{P_2\},\quad\{\mathcal{I}\},\quad
\{(0,0),(1,0)\},\quad\{(1,0),P_1\},\quad\{P_1,P_2\}\,.\] In the
above $\mathcal{I}$ means interior equilibrium. We observe that if
$r_{12}=r_{11}$, the pattern $\{P_1, P_2\}$ (Lemma 3.3.C.1(c) in
Table 2) did not show up in our simulations. We now prove that this
pattern cannot occur. From Table 2, sixth line from the bottom, we see that
$S_1, S_2$ are in the superior case ($r_{33} < r_{22} < r_{23},\,
r_{33} < r_{11} < r_{13}$). This and the fact $r_{12}=r_{11}$ imply
that the signs of the Greek alphabets defined in (\ref{greekdef})
are
\[
\alpha_1 \;(0),\quad \alpha_2 \;(+),\quad \alpha_3\;(+),\quad
\beta_1 \;(-),\quad \beta_2 \;(+), \quad\beta_3\; (?),\quad
\gamma_1\; (?),\quad \gamma_2 \;(-),\quad \gamma_3\; (-)\,.
\]
The signs of the eigenvalues of $J$ at
$P_1, P_2$ are $P_1\{-,\frac{-U_1}{-}\}$ and
$P_2\{-,\frac{-U_2}{-}\}$. If $P_1, P_2$ are both stable, we have
$U_1 < 0$ and $U_2 < 0$. From (\ref{Uformula}), $U_1 = \gamma_2
\beta_1 - \gamma_1 \beta_2 = (-)(-) - \gamma_1 (+)$ and $U_2 =
\alpha_2 \gamma_1 - \alpha_1 \gamma_2 = (+)\gamma_1$ which imply
that $U_1, U_2$ cannot both be negative. Therefore, the pattern
$\{P_1, P_2\}$ cannot occur. If $r_{12}=r_{22}$, then among the
above $7$ patterns, only the pattern $\{\mathcal{I}\}$ (Lemma
3.3.C.1(b) in Table 1) cannot occur. The proof of this fact is
similar and will be omitted.
\smallskip

(b) Suppose $S_2$ is completely dominant ($r_{13}=r_{11}$ or
$r_{13}=r_{33})$.\,\,We repeat the process above and find that there
are $19$ cases of $S_2$ being in the intermediate case which
generate $7$ patterns:
\[\{(1,0)\},\quad\{P_1\},\quad \{P_3\},\quad \{\mathcal{I}\},\quad\{(0,1),
(1,0)\},\quad \{(1,0), P_1\},\quad\{P_1, P_3\}\,.\] However, if
$r_{13} = r_{11}$, the pattern $\{P_1,P_3\}$ (Lemma 3.3.B.1(c) in
table 2) cannot occur. If $r_{13} = r_{33}$, the pattern
$\{\mathcal{I}\}$ (Lemma 3.3.B.1(b) in table 1) cannot occur. The
proof of these fact will be omitted.\smallskip

(c) Suppose $S_1$ is completely dominant ($r_{23}=r_{22}$ or
$r_{23}=r_{33})$.\,\,We repeat the process above and find that there
are $17$ cases of $S_1$ being in the intermediate case which
generate $7$ patterns:
\[\{(1,0)\},\quad\{P_2\},\quad\{P_3\},\quad\{\mathcal{I}\},\quad\{(0,1), (1,0)\},\quad\{(0,1),
P_2\},\quad\{P_2, P_3\}\,.\] However, if $r_{23} = r_{22}$, the
pattern $\{P_2,P_3\}$ (Lemma 3.3.A.1(c) in table 2) cannot occur and
if $r_{23} = r_{33}$, the pattern $\{\mathcal{I}\}$ (Lemma
3.3.A.1(b) in table 1) cannot occur. The proof of these facts will
also be omitted.\medskip

\noindent (ii)\,\,Complete dominance on two sides.\smallskip

There are $12$ ways to assume complete dominance on two sides. The
complete list is given below. They are all confirmed by computer
simulations.\smallskip

(a)  $S_3$ and $S_2$ are completely dominant:
$\{(1,0)\},\;\{(1,0),P_1\}$\,.\smallskip

(b)  $S_3$ and $S_1$ are completely dominant:
$\{(1,0)\},\;\{P_2\}$\,.\smallskip

(c)  $S_2$ and $S_1$ are completely dominant:
$\{(1,0)\},\;\{P_3\},\;\{(0,1),(1,0)\}$\,.\medskip

\noindent (iii)\,\,Complete dominance on all sides.\smallskip

There are $8$ ways to assume complete dominance on all sides. They
all generate the same pattern $\{(1,0)\}$.
\smallskip

%
%

\section{Higher Number of Alleles}\label{higherallele}

For $k$ alleles, $k\geq 4$, there are too many possible patterns
that we cannot prove their existence like we did for the
three-allele model in Section~\ref{noncompletedominance}. We therefore
resort to computer simulations to verify that certain patterns exist.
We also develop some rules to show that certain patterns do not exist.
For example, according to Proposition~\ref{interiorstable}, we have\medskip

\noindent {\it Rule 1}.\,\, For any $k$, if the interior equilibrium
is stable, then no other equilibrium can be stable.\medskip

We say that three edge equilibria are \emph{coplanar} if the edges
they lie on form a triangle. From the results of the three-allele
model, we have the following\medskip

\noindent {\it Rule 2}.\,\, The maximum number of coexisting stable
equilibria in $\Delta$ is $3$ and they must be the \vspace{.01in}

\hspace{.43in}three vertices. Hence, stable coplanar edge equilibria
is not possible.
\medskip

\noindent {\it Rule 3}.\,\, If the existing stable equilibria are
all vertices, then they must include the vertex $(1,0)$.\medskip

Rule 3 is a consequence of condition (\ref{riicond}). We
conjecture that it is valid for any $k$.

\subsection{Four-allele Model}\label{4allele}
The invariant set $\Delta$ defined by (\ref{deltadef}) is a
tetrahedron. It contains $15$ possible equilibria including $4$
vertices, $6$ edge equilibria, $4$ face equilibria, and one interior
equilibrium. From the results of the three allele case, it is easy
to see that the maximum number of coexisting stable equilibria is
$4$ which we state below as rule 4. We first construct a table of
all $64$ possible patterns and eliminate those that cannot exist
using rules 1, 2 and 6 below. For example, the pattern $3v+f$ (three
stable vertices and one stable face equilibrium) is not possible
because such a pattern would imply a three-allele model with two
stable vertices and a stable edge equilibrium which violates rule 2.
We also eliminate the patterns $3f$, $e+2f$ and $3e+f$ which are
shown to be not possible in \cite{vickers1}. The remaining patterns
are shown to exist by computer simulations and are summarized in
Table \ref{tab4} below. In the table, $N_{s}$ denotes the number of
coexisting stable equilibria and $N_{p}$ denotes the number of
patterns for the type indicated. If we add up all the numbers in the
$N_p$ column, we see that the four-allele model has $117$ patterns.
The last column in the table explains how the numbers in the column
$N_p$ are obtained.  For example, according to Table~\ref{tab4},
there are $12$ patterns with one vertex equilibrium and one edge equilibrium; that is, $1v +
1e$. There are six edges in a tetrahedron and we can pair anyone
(hence ${6\choose 1} = 6$ choices) with any one of the four vertices
that does not lie in the edge. Since a tetrahedron has four vertices
and each edge connects two vertices, the number of choices here is
${4-2\choose 1} = 2$. The notation $m\choose n$ means $m$ choose $n$
which equals $m!/(n!(m-n)!)$. Therefore, there are $6\times 2 = 12$
such patterns. We develop the following rules which will be used in
the study of the five-allele model. The proof of rule 6 is rather
involved and will not be presented here. Rule 5 follows from looking
at Table 4.
\medskip

\noindent {\it Rule 4}.\,\, The maximum number of coexisting stable
equilibria in $\Delta$ is $4$.\medskip

\noindent {\it Rule 5}.\,\, If $\Delta$ contains three or more stable
equilibria, then none can be on a face except we can
have a face with a stable equilibrium and two stable
edge equilibria not on the face.\medskip

\noindent {\it Rule 6}.\,\, If the stable equilibria are all
vertices, then they must include the vertex $(1,0,0)$.\medskip

\begin{table}[H]
\centering
\begin{tabular}{|l|l|l|l|l|r|l|l|}
\hline
 \multirow{2}{*}{$N_{s}$}& \multicolumn{4}{|c|}{pattern type}
& \multirow{2}{*}{$N_{p}$} & \multirow{2}{*}{\hspace{1.25in}Explanations} \\
\cline{2-5} & $v$ & $e$ & $f$ & $t$ & & \\ \hline \multirow{4}{*}{1}
& 1 & & &  & 1 & must be $(1,0,0)$. \\ \cline{2-7} &  & 1 &  &  & 6
&
any one of the six edges.\\ \cline{2-7} &  &  & 1 &  & 4 & any one of the four faces.\\
\cline{2-7} & & &  & 1 & 1 & the interior equilibrium. \\ \hline
\multirow{7}{*}{2} & 2 &  &  &  & 3 & any two vertices one of
which must be $(1,0,0)$.\\
\cline{2-7} &  & 2 &  &  & 15 & any two edges: ${6\choose 2}=15$. \\
\cline{2-7} &  &  & 2 &  & 6 & any two faces: ${4\choose 2}=6$. \\
\cline{2-7} & 1 & 1 &  &  & 12 & pair any edge with a vertex not in
the edge: ${6\choose 1}\times {4-2\choose 1}=12$. \\ \cline{2-7} & 1
&  & 1 &  & 4 & pair any face with the one vertex not in the face.\\
\cline{2-7} &  & 1 & 1 &  & 12 & pair any face with an edge not in
the face: ${4\choose 1}\times{6-3\choose 1}=12$.
\\ \hline \multirow{4}{*}{3} & 3 &  &  &  & 3 & three
vertices one of which must be $(1,0,0)$: ${4-1\choose 2}=3$. \\
\cline{2-7} & & 3 &  &  & 16 & any three edges not coplanar:
${6\choose 3}-4=16$.
\\ \cline{2-7} & 2 & 1 &  &  & 6 & pair any edge with the two
vertices not in the edge.\\ \cline{2-7}
& 1 & 2 &  &  & 12 & pair any vertex with two edges not containing the vertex:\\
&  &  &  &  &  & ${4\choose 1}\times {3\choose 2}=12$. Note two ends of an edge are vertices. \\
\cline{2-7}
& & 2 & 1 &  & 12 & any face with two of the three edges not from the face.\\
\hline
\multirow{2}{*}{4} & 4 &  &  &  & 1 & all four vertices.\\
\cline{2-7} &  & 4 &  &  & 3 & any four edges without forming a
triangle: ${6\choose 4}-4\times 3=3$.\\ \hline
\end{tabular}
\vspace*{.1in}
\caption{Four-allele model with 117 patterns. In the table $v$ denotes vertices, $e$
denotes edge equilibria, $f$ denotes face equilibria,
and $t$ denotes tetrahedral equilibria.}\label{tab4}
\end{table}

\subsection{Five-allele Model}\label{5allele}

The set $\Delta$ contains $31$ equilibria including $5$ vertices,
$10$ edge equilibria, $10$ face equilibria, $5$ tetrahedral
equilibria and $1$ interior equilibrium denoted by $i$. There are
at least $2351$ patterns shown in the tables below. 
There are a few patterns which we cannot show nonexistence nor found them in our simulations.
They are discussed at the end of this subsection.

\begin{table}[H]
\centering
\begin{tabular}{|l|l|l|l|l|l|r|l|l}
\hline
\multirow{2}{*}{$N_{s}$}& \multicolumn{5}{|c|}{pattern type}
& \multirow{2}{*}{$N_{p}$} & \multirow{2}{*}{\hspace{1.25in}Explanations} \\
\cline{2-6} & $v$ & $e$ & $f$ & $t$ & $i$ & & \\ \hline
\multirow{5}{*}{1} & 1 &  &  &  & & 1 & must be $(1,0,0,0).$ \\
\cline{2-8} &  & 1 &  &  & &
10 & any one of the ten edges.\\ \cline{2-8} &  &  & 1 &  &  & 10 & any one of the ten faces.\\
\cline{2-8} &  &  &  & 1 &  & 5 & any one of the five tetrahedrons.
\\ \cline{2-8} &  &  &  &  & 1 & 1 & the interior equilibrium.
\\ \hline \multirow{10}{*}{2} & 2 &  & & & & 4 & one of
the two vertices must be $(1,0,0,0)$.\\
\cline{2-8} &  & 2 &  &  &  & 45 & any two edges: ${10\choose
2}=45.$
\\ \cline{2-8} &  &  & 2 &  &  & 45 & any two faces: ${10\choose
2}=45.$ \\ \cline{2-8} &  &  &  & 2 &  & 10 & any two tetrahedrons:
${5\choose 2}=10.$ \\ \cline{2-8} & 1 & 1 &  &  &  & 30 & any edge
and a vertex not in the edge: $10 \times 3=30$. \\ \cline{2-8} & 1 &
& 1 &  &  & 20 & any face and a
vertex not in the face: $10 \times 2=20$. \\
\cline{2-8}
& 1 &  &  & 1 &  & 5 & tetrahedron and the vertex not in the tetrahedron.\\
\cline{2-8} &  & 1 & 1 &  &  & 70 & any face with
an edge not in the face: $10 \times 7=70.$ \\
\cline{2-8}
&  & 1 &  & 1 &  & 20 & any tetrahedron and an edge not in the tetrahedron:\\
& & & & & & & ${5\choose 1}\times {10-6\choose 1}=20$. \\
\cline{2-8}
&  &  & 1 & 1 &  & 30 & any tetrahedron and a face not in the tetrahedron:\\
& & & & & & & ${5\choose 1}\times {10-4\choose 1}=30$. \\ \hline
\multirow{10}{*}{3} & 3 &  &  &  &  & 6 & one of the three vertices
must be $(1,0,0,0)$: ${5-1\choose 2}=6$. \\ \cline{2-8} &  & 3 &  &
& & 110 & any three edges not coplanar: ${10\choose 3}-10=110$.
\\ \cline{2-8}
&  &  & 3 &  &  & 100 & any three faces not in a common tetrahedron: \\
& & & & & & & ${10\choose 3}-{4\choose 3}\times 5=100$. \\
\cline{2-8} & 2 & 1 &  &  &  & 30 & pair any edge with two vertices
not in the edge: \\ & & & & & & & ${10\choose 1}\times {5-2\choose
2}=30$. \\ \cline{2-8} & 2 &  & 1 &  &  & 10 & pair any face with
the two vertices not in the face.\\ \cline{2-8} & 1 & 2 & & & & 75 &
any vertex with two edges not containing the
vertex: \\ & & & & & & & ${5\choose 1}\times {10-4\choose 2}=75$.\\
\cline{2-8} &  & 2 & 1 &  &  & 210 & pair any face with two edges
not in the face: \\ & & & & & & & $10\times {10-3\choose 2}=210 $.
\\ \cline{2-8}
&  & 2 &  & 1 &  & 30 & pair any tetrahedron with any
two edges not in the \\ & & & & & & & tetrahedron: ${5\choose
1}\times{10-6\choose
2}=30$.\\
\cline{2-8} &  &  & 2 & 1 &  & 75  & pair any tetrahedron with any
two faces not in the \\ & & & & & & & tetrahedron: ${5\choose
1}\times{10-4\choose 2}=75$.\\ \cline{2-8}
&  & 1 & 2 &  &  & 180 & any edge with two faces not containing the edge and \\
& & & & & & & not in a common tetrahedron with the edge (an edge\\
& & & & & & & is shared by three faces and three tetrahedrons): \\
& & & & & & & $10\times({10-3\choose 2}-3)=180$.\\ \hline
\end{tabular}
\vspace*{.1in}
\caption{ Five-allele model with at least 2351 patterns.}\label{tab5}
\end{table}

\bigskip
\begin{table}[H]
\centering
\begin{tabular}{|l|l|l|l|l|l|r|l|l|}
 \hline \multirow{2}{*}{$N_{s}$}&
\multicolumn{5}{|c|}{pattern type} & \multirow{2}{*}{$N_{p}$} &
\multirow{2}{*}{\hspace{1.25in}Explanations} \\ \cline{2-6} & $v$ & $e$ & $f$ & $t$ & $i$ & & \\
\hline \multirow{3}{*}{3}& 1 &  & 2 &  &  & 30 & pair any vertex
with two faces not containing the vertex \\ & & & & & & & (a vertex
is shared by six faces): ${5\choose 1}\times{10-6\choose 2}=30$. \\
\cline{2-8} & 1 & 1 & 1 &  &  & 60 & a face with a vertex not
in the face and an edge neither \\ & & & & & & & in the face nor
containing the vertex: $10\times 2\times 3=60$.\\
\cline{2-8} &  & 1 & 1 & 1 &  & 60 & any tetrahedron with a face not
in the tetrahedron plus \\ & & & & & & & an edge neither in the
tetrahedron nor in the face:\\ & & & & & & & $5\times 6 \times
2=60$.
\\ \hline
\multirow{12}{*}{4} & 4 &  &  &  &  & 4 & one of the four vertices
must be $(1,0,0,0)$: ${5-1\choose 3}=4$.  \\ \cline{2-8} & & 4 &  &
& & 140 & any four edges without forming a triangle:\\ & & & & & & &
${10\choose 4}-{10\choose 1}\times{10-3\choose 1}=140$.
\\ \cline{2-8} &  &  & 4 &  &  & 75 & see A below.
\\ \cline{2-8} & 2 & 2 &  &  &  & 30 & pair any two vertices with
any two edges not containing \\ & & & & & & & these vertices:
${5\choose 2}\times{3\choose 2}=30$.  \\ \cline{2-8} &  & 2 & 2 &  &
& 150 & see B below.\\ \cline{2-8} & 3 & 1 &  &  &  & 10 & pair any
edge with the three vertices not in the edge.
\\ \cline{2-8} & 1 & 3 &  &  &  & 80 & any three edges without
forming a triangle from a \\ & & & & & & & tetrahedron (see pattern
3e in the four-allele model) \\& & & & & & &
and the vertex not in the tetrahedron: ${16\times 5}=80$.\\
\cline{2-8} &  & 3 & 1 &  &  & 60 & see C below.\\
\cline{2-8} &  & 3 &  & 1 &  & 20 & pair any tetrahedron with any
three edges not in the \\ & & & & & & & tetrahedron: ${5\choose
1}\times{4\choose 3}=20$. \\ \cline{2-8} &  & 1 & 3 &  &  & 120 &
see
D below.\\ \cline{2-8} &  &  & 3 & 1 &  & 60 & see E below. \\
\cline{2-8} & 1 & 2 & 1 &  &  & 60 & pick two edges and one face
from a tetrahedron and \\ & & & & & & & choose the vertex not in the
tetrahedron: $5\times 12=60$.\\ & & & & & & & (see pattern $v+2e$ in
four-allele model).\\\hline \multirow{6}{*}{5} & 5 & &  &  & & 1 &
all five vertices.
\\ \cline{2-8} &  & 5 &  &  &  & 72 &
see F below.
\\ \cline{2-8} &  &  & 5 &  &  & 12 & see G below.
\\ \cline{2-8} & 1 & 4 &  &  &  & 15 & any four edges
without forming a triangle from a \\ & & & & & & & tetrahedron (see
pattern 4e in the four-allele model) \\ & & & & & & & and the vertex
not in the tetrahedron: $3\times 5=15$.\\ \cline{2-8} &  & 4 & 1 & &
& 90 & type face $\{123\}$ with edges $\{14\}$ $\{15\}$ $\{24\}$
$\{25\}$: $10\times 3$.\\ & & & & & & & type face $\{123\}$ with
edges $\{14\}$ $\{35\}$ $\{24\}$ $\{45\}$: $10\times 6$.\\
\cline{2-8} &  & 3 & 2 &  &  & 60 & see H below.
\\ \hline 6 &  & 6 &  & &  & 10 & choose the six edges in any two tetrahedrons not on
\\ & & & & & & &  the common face of the two tetrahedrons:
${5\choose 2}=10$.\\ \hline
\end{tabular}
\vspace*{.1in}
\caption{ Five-allele model with at least 2351 patterns
(continued).}\label{tab6}
\end{table}

To understand the explanation column of the above table, take the
first pattern $1v+2f$ in Table~\ref{tab6}. There are $5\choose 1$
vertices and suppose a vertex, say vertex $1$, is chosen. Then among
the $10$ faces of $\Delta$, ${4\choose 2} = 6$ of them contains
vertex $1$. Therefore, $10-6$ faces do not contain vertex $1$ and
there are ${10-6\choose 2} = 6$ ways to choose two of them. This
explains why $N_p = 5\times 6 = 30$. The explanations of some the
patterns in Table~\ref{tab6} are too complicated to fit into the
space provided in the table and are explained below.\medskip

A.\quad For the $4f$ pattern, we observe that each face is shared by
two tetrahedrons so there are actually $8$ faces if we take the
tetrahedrons into account. According to Rule 5, each tetrahedron can
contain at most 2 stable face equilibria. The $8$ faces may be put
into the $5$ tetrahedrons in two ways: $2+2+2+2+0$ or $2+2+2+1+1$.
Consider the first way. There are ${5 \choose 4} \times 3 = 15$
patterns. The number $3$ may be explained as follows. Suppose we
label the vertices $1$ through $5$ and we eliminate the tetrahedron
$\{2345\}$. Consider one of the remaining four tetrahedrons, say
$\{1234\}$. There are ${3 \choose 2} = 3$ ways to choose $2$ faces
from $3$ because the face $\{234\}$, being on $\{2345\}$, is not
eligible. But once we choose two of the faces, the other two faces
on each of the remaining $3$ tetrahedrons are fixed. This explains
the $3$ in the formula. For the second way, there are ${5\choose
2}\times 6=60$ patterns so altogether we have $15+60=75$ $\;\;4f$
patterns.\medskip

B.\quad For the $2e+2f$ pattern, it is easy to see that the two
faces chosen must have one common edge ($10\times {3\choose 2}=30$
ways) or one common vertex ($5\times {4\choose 2}/2=15$ ways because
of symmetry). In each case, there are 6 ways to choose the two
stable edge equilibria in order not to violate Rule 5. Thus,
$30\times 6 + 15\times 6 =270$\ patterns are expected. However, the
pattern with faces $\{123\}$, $\{134\}$ and edges $\{15\}$, $\{25\}$
is not show up in our simulations and we could not find it in any
literature. There are $30\times 4=120$ such patterns. Therefore,
there are at least $150$ and at most $270$\, $2e+2f$\,
patterns.\medskip

C.\quad For the $3e+1f$ pattern, first pick one of the $10$ faces,
then choose $3$ edges such that they are not coplanar (Rule 2), none
of them is from the face chosen (Rule 1), and they and the face are
not in a common tetrahedron (Rule 5). Accordingly, there are
$10\times ({7\choose 3}-3-2)=300$ patterns expected. But only
patterns of the type face $\{123\}$ with edges $\{14\}$, $\{25\}$,
$\{45\}$ ($10\times 3\times 2=60$ patterns) show up in our
simulations, whereas the patterns of the type face $\{123\}$ with
edges $\{14\}$, $\{34\}$, $\{45\}$ ($60$ patterns), of the type face
$\{123\}$ with edges $\{14\}$, $\{34\}$, $\{15\}$ ($60$ patterns),
and of the type face $\{123\}$ with edges $\{14\}$, $\{34\}$,
$\{25\}$ ($60$ patterns) do not show up in our simulations.
Therefore, there are at least $60$ and at most $300$ 3e+1f\,
patterns.\medskip

D.\quad For the $1e+3f$ pattern, there are two types of patterns.
For type 1, first pick one of the $10$ faces, say, $\{345\}$ as face
$1$, then face 2 is determined by vertex $\{1\}$ and any edge of
face $1$, say, $\{134\}$, face $3$ is determined by vertex $\{2\}$
and any edge of face $1$ but not used by face $2$, say, $\{245\}$.
The stable edge equilibrium is on $\{12\}$. There are $10\times
3\times 2=60$ patterns of type 1. For type 2, first pick two of the
$10$ faces with a common edge ($30$ ways), say, $\{134\}$ and
$\{234\}$, then the third face is $\{125\}$ decided by the left
vertex $\{5\}$ and the two vertices not on the common edge of the
first two faces. Lastly, the stable edge equilibrium could be either
on $\{35\}$ or $\{45\}$. There are $30\times 2=60$ patterns of type
2. \medskip

E.\quad For the $3f+1t$ pattern, first pick any tetrahedron, say,
$\{1234\}$. To decide the three faces, we pick three edges from
$\{1234\}$ which are not coplanar and not all from the same vertex,
say, $\{12\}$, $\{13\}$, $\{34\}$, then the three faces are
determined by vertex $\{5\}$ and these three edges respectively,
namely, $\{125\}$, $\{135\}$, $\{345\}$. There are total $5\times
({6\choose 3}-4-4)=60$ patterns.\medskip

F.\quad For the $5e$ pattern, we observe that there are $252$ ways
to choose $5$ edges from the $10$ edges but we must subtract $150$
cases of one coplanar edges and $30$ cases of two coplanar edges
which will result in only $72$\, 5e\, patterns. To see that there
are $150$ one coplanar edges, suppose the coplanar edges are the
sides of the triangle $\{345\}$ (we have $10$ such triangles). Then
the remaining two edges can be chosen two ways (so that we don't
have two coplanar edges). One way is to choose the edge $\{12\}$ and
then the fifth edge may be chosen from any of the six edges
$\{13\},\{14\},\{15\},\{23\},\{24\}$,$\{25\}$. A second way is to
choose one edge from the group $\{13\},\{14\},\{15\}$ and another
edge from the group $\{23\},\{24\},\{25\}$ with a total of $9$
choices. Hence, there are $6 + 9 = 15$ ways to choose the other two
edges besides the sides of the triangle $\{345\}$. Therefore, there
are $150$ cases of one coplanar edges. The argument for the $30$ two
coplanar edges is similar and will be omitted.\medskip

G.\quad For the $5f$ pattern, each face is shared by two
tetrahedrons so there are actually $10$ faces if we take the
tetrahedrons into account and each tetrahedron has exactly two
stable face equilibria. Choose two faces (6 ways) from the first
tetrahedron $\{1234\}$, say, $\{123\}$ and $\{124\}$. Then in the
second tetrahedron $\{1235\}$, we choose one more face either
$\{235\}$ or $\{135\}$. The two faces chosen in the rest
tetrahedrons are fixed now. Thus, there are $6\times 2=12$
ways.\medskip

H.\quad For the $3e+2f$ pattern, choose any two faces with a common
edge (30 ways), say, $\{234\}$ and $\{345\}$, then choose three
edges out of $\{12\},\{13\},\{14\},\{15\}$ such that they are not in
a common tetrahedron with either $\{234\}$ or $\{345\}$ by Rule 5.
Thus, there are $30\times 2=60$ ways.\bigskip

We cannot find the following patterns in our simulations nor can we prove their nonexistence: $1f+2t$, $2f+2t$, $1e+2f+1t$,
$4f+1t$, some types of $2e+2f$ (see B above), and some types of
$3e+f$ (see C above).  From the tables we observe the following
rules which may be used to study the six-allele model:\medskip

\noindent {\it Rule 7}.\,\, The maximum number of coexisting stable
equilibria for the five-allele model is $6$ with\smallskip

\hspace{.4in} all stable equilibria on the edges.\medskip

\noindent {\it Rule 8}.\,\, If there are five stable equilibria, none of them can be in the interior of a tetrahedron.\medskip

\noindent {\it Rule 10}.\,\, If the stable equilibria are all
vertices in five-allele case, then one of them must be\smallskip

\hspace{.4in} $(1,0,0,0)$.

\subsection{Computer Simulations}
\label{simulate} \setcounter{equation}{0} In this section we briefly
describe how one can perform computer simulations to help find
patterns for higher number of alleles. One possible idea which we
adopted is the following. First write a computer program that
randomly generates a fitness matrix $R$ that satisfies condition
(\ref{riicond}). Then use the method described in
Remark~\ref{remark1} to compute and count the number of negative
eigenvalues at each existing equilibrium.  Note that for $k$
alleles, there are a maximum possible $2^k-1$ equilibria. Write the
results of the computer program to a text file.  Each line in the
text file should contain $2^k-1$ integers representing the number of
negative eigenvalues of the Jacobian matrix evaluated at that
particular equilibrium. A special symbol should be used in case the
equilibrium does not exist. Run a sorting algorithm on the text file
that will turn each line into a pattern (see example below). One
needs to repeat this process several times, each time with
increasing number of simulations until no new patterns are found.
\smallskip

As an example, suppose we want to find all the patterns for the
three-allele model. Then a line in our output text file may look
like $-1 -2 -2 +0\, +9 -1 +9$, which means that the three vertices
have $1, 2$ and $2$ negative eigenvalues, respectively, $P_1, P_3$
have $0$ and $1$ negative eigenvalues, respectively, and $P_2$ and
the interior equilibrium do not exist for this round of simulation
($9$ means the corresponding equilibrium does not exist). Of course
another fitness matrix $R$ will generate a different line of
numbers. The sorting algorithm will turn this line into the pattern
$\# -2 -2 \;\# \;\# \;\# \;\# \;\#$ and write it to a file called
"filename.2" which contains all distinct patterns with two stable
equilibria. ``filename'' is a user supplied name and the sorting
algorithm also removes duplicate patterns. At the end, the sorting
algorithm produces three files: "filename.1", "filename.2" and
"filename.3". "filename.1" contains $30$ distinct patterns with one
stable equilibrium (see Table~\ref{tab1}), "filename.2" contains
$35$ distinct patterns with two stable equilibria (see
Table~\ref{tab2}) and "filename.3" contains $4$ distinct patterns
with three stable equilibria (see Table~\ref{tab3}).\smallskip

Some patterns are difficult to find if $k$ is large. For example,
for the five-allele case, we need to simulate over $40$ million
times until no new patterns are found and yet we are not sure if we
found them all. For the six-allele case, we used $20$ processors and
simulated over $400$ million times and found over $60,000$ distinct
patterns. Of course, not finding a pattern does not mean that
pattern does not exist unless its absence can be proved
theoretically. We believe that the number of patterns increases
exponentially with $k$.
\smallskip

\section{Conclusion}
\label{conclusion} \setcounter{equation}{0} In this paper, we list
all patterns for the three- and four-allele selection models and
almost all patterns for the five-allele selection model for all
fitness matrices satisfying condition (\ref{riicond}).  A pattern is
simply the number and locations in the invariant set $\Delta$ of all
coexisting stable equilibria. In our investigation, we do not assume
the truth of the conjecture of Cannings and Vickers that any subset
of a pattern is also a pattern \cite{vickers1,vickers4}. Knowing
the patterns is important because it tells us the possible long-term
genetic makeup of the population; that is, which alleles will
survive and which will disappear under selection forces in the long
run. For the three-allele model, we prove our results analytically
and show the separatrices for each pattern so that we know the
asymptotic behavior of the solutions of the mathematical model
(\ref{basicmodel}) as $t \rightarrow \infty$. We also consider the
case of complete dominance for the three-allele model. Through the
use of computer simulations and the rules we developed, we show that
there are $117$ patterns for the four-allele model and at least
$2351$ patterns for the five-allele model. There are a few patterns
in the five-allele model that did not show up in our computer
simulations but we are not able to show that they do not exist. For
the six-allele model, parallel computer simulations indicate that
there are over $60,000$ patterns. We conjecture that the number of
patterns grow exponentially as the number of alleles increases.
In this paper, we also prove several interesting mathematical
results in Section~\ref{generic} some of which we believe are new.

\section{Acknowledgments}  We thank Allan E. Johannesen
of the WPI Computing and Communications Center for writing the
sorting algorithm to help us recognize patterns. We also thank
Siamak Najafi and Adriana Hera for their help with parallel
computing for the six-allele case. Finally, we thank Jesse Bowers
for helpful discussions on Lemmas~\ref{noneexist} - \ref{allexist}.



\end{document}